\documentclass[prd,twocolumn,showpacs,floats,floatfix,nofootinbib]{revtex4}

\usepackage{graphicx}
\usepackage{epstopdf}
\usepackage{color}
\usepackage{natbib}
\usepackage{dcolumn}
\usepackage{amssymb}
\usepackage{bm}
\usepackage{amsfonts}
\usepackage{txfonts}
\usepackage{bbm}
\usepackage{bbold}

\usepackage[pdftex,left=1in,right=1in,top=1in,bottom=1in]{geometry}

\def\spose#1{\hbox to 0pt{#1\hss}}

\def\lta{\mathrel{\spose{\lower 3pt\hbox{$\mathchar"218$}}
     \raise 2.0pt\hbox{$\mathchar"13C$}}}
\def\gta{\mathrel{\spose{\lower 3pt\hbox{$\mathchar"218$}}
     \raise 2.0pt\hbox{$\mathchar"13E$}}}
\newcommand{\ex}{\mathrm{e}}
\newcommand{\dd}{\mathrm{d}}

\newcommand{\ie}{\textsl{i.e.~}}
\newcommand{\cf}{\textsl{cf.~}}
\newcommand{\eg}{\textsl{e.g.~}}
\newcommand{\etal}{\textsl{et al.~}}

 \newcommand{\cs}{{\Upsilon}}
\newcommand{\GN}{G_{_\mathrm{N}}}

\newcommand{\Mp}{M_{_\mathrm{Pl}}}

\def\beq{\begin{equation}}
\def\eeq{\end{equation}}
\def\bea{\begin{eqnarray}}
\def\eea{\end{eqnarray}}
\def\eqref{\ref}
\def\x{\mathrm{x}}
\def\y{\mathrm{y}}
\def\n{\mathrm{n}}
\def\s{\mathrm{s}}

\def\Bx{{\cal B}^\x}

\def\Axy{{\cal A}^{\x \y}}
\def\LX{{\cal X}^\mu}
\def\LY{{\cal Y}^\mu}

\def\nx{n_\x}
\def\ny{n_\y}
\def\nxy{n_{\x \y}}
\def\utx{u^t_\x}
\def\uzx{u^z_\x}
\def\uty{u^t_\y}
\def\uzy{u^z_\y}

\def\vx{V_\x}

\def\vn{V_\n}
\def\vs{V_\s}
\def\Nx{{\cal N}_\x}
\def\Mx{{\cal M}_\x}

\def\Nn{{\cal N}_\n}
\def\Mn{{\cal M}_\n}
\def\Ns{{\cal N}_\s}
\def\Ms{{\cal M}_\s}

\def\cx{c_{\x}}

\def\cn{c_\n}
\def\cs{c_\s}
\def\Cxy{\mathcal{C}_{\x \y}}
\def\Cyx{\mathcal{C}_{\y \x}}
\def\Cns{\mathcal{C}_{\n \s}}
\def\Csn{\mathcal{C}_{\s \n}}

\def\setR{\mathbb{R}}
\def\setC{\mathbb{C}}

\begin{document}
\title{Multi-fluid cosmology: An illustration of fundamental principles}

\author{G.~L.~Comer}
\email{comergl@slu.edu}
\affiliation{Department of Physics \& Center for Fluids at All Scales, Saint
Louis University, St.~Louis, MO, 63156-0907, USA}

\author{Patrick Peter}
\email{peter@iap.fr}
\affiliation{${\cal G}\setR\varepsilon\setC{\cal O}$ -- Institut
d'Astrophysique de Paris, UMR7095 CNRS, Universit\'e Pierre \& Marie Curie,
98 bis boulevard Arago, 75014 Paris, France}

\author{N.~Andersson}
\email{N.A.Andersson@soton.ac.uk}
\affiliation{School of Mathematics, University of Southampton,
Southampton SO17 1BJ, UK}

\date{\today}

\begin{abstract}
  Our current understanding of the Universe depends on the interplay
  of several distinct ``matter'' components, which interact mainly
  through gravity, and electromagnetic radiation. The nature of the 
  different components, and possible interactions, tends to be based on 
  the notion of coupled perfect fluids (or scalar fields). This approach is 
  somewhat naive, especially if one wants to be able to consider issues 
  involving heat flow, dissipative mechanisms, or Bose-Einstein condensation 
  of dark matter. We argue that a more natural starting point would be the  
  multi-purpose  variational relativistic multi-fluid system that has so far 
  mainly been applied to neutron star astrophysics. As an illustration of the 
  fundamental principles involved, we develop the formalism for determining 
  the non-linear cosmological solutions to the Einstein equations for a general 
  relativistic two-fluid model for a coupled system of matter (non-zero 
  rest mass) and ``radiation'' (zero rest mass). The two fluids are allowed to 
  interpenetrate and exhibit a relative flow with respect to each other, implying, 
  in general, an anisotropic Universe. We use initial conditions such that the 
  massless fluid flux dominates early on so that the situation is effectively that    
  of a single fluid and one has the usual 
  Friedmann-Lema\^{\i}tre-Robertson-Walker (FLRW) spacetime. We find that 
  there is a Bianchi I transition epoch out of which the matter flux dominates.   
  The situation is then effectively that of a single fluid and the spacetime 
  evolves towards the FLRW form. Such a transition opens up the possiblity of 
  imprinting observable consequences at the specific scale corresponding to 
  the transition time. 
\end{abstract}


\maketitle
	
\section{Introduction}

The cosmological principle states that the Universe is homogeneous and
isotropic. Given the increased quality of cosmological observations, this 
fundamental principle is now becoming testable, and indeed questionable. 
That questions abound in this area is obvious from the fact that we do not 
have a good handle on the nature of dark components that dominate the 
cosmological ``standard model'' \cite{PPJPU}.  A large number of alternative 
models and theories have been suggested in the literature, but most are not 
particularly compelling. The treatment of the different matter components, in 
particular, is often based on the notion of coupled perfect fluids or scalar 
fields. If we are to understand the bigger picture, we need to make progress 
on this aspect, especially if we want to be able to consider issues like heat flow 
\cite{modak,pavon,AL11}, dissipative mechanisms 
\cite{Weinberg71,PK91,Velten:2011bg}, Bose-Einstein condensation 
of dark matter \cite{SY09,Harko11} and possibly many others.

We argue that a more natural starting point for this endeavor would be the 
relativistic variational multi-fluid approach \cite{carter89:_covar_theor_conduc} 
that has (so far) mainly been applied to neutron star astrophysics 
\cite{andersson07:_livrev}, and recently to relativistic beams and shocks 
\cite{NBM11}. This approach would seem natural since there could have been 
phases during which the Universe would have effectively been anisotropic, with 
different components evolving ``independently''. For the most part, models 
discussed in the current literature, including initially anisotropic geometries, 
describe the matter content in terms of either effectively many component 
single fluid models \cite{Gromov:2002ek}, or a plain single component 
\cite{EGcCP07,PPU08,KM10}; although isotropisation is expected in such 
situations, as required to end up with a realistic (read: in agreement with 
currently available data) model \cite{DLH09}, interesting new consequences 
can however be derived, \eg by enhancing an initially vanishingly small 
non-gaussian signal \cite{DP11}.

As an illustration of the fundamental principles involved, we develop the
formalism for determining the cosmological solutions to the Einstein
equations for a general relativistic, two-fluid model coupling matter (non-zero 
rest mass) and ``radiation'' (zero rest mass). Drawing on the experience from 
other applications it would be straightforward to consider other relevant cases, 
\eg involving a dissipative heat flow \cite{AL11} or superfluid condensates  
\cite{SY09,Harko11}. However, the chosen example is perhaps the most 
conventional, since the leading-order thermodynamics of massless particles 
has some generic features (compare, say, a photon and phonon gas), and the 
same for a massive component when the density becomes (relatively) small. 

Within this context, we will demonstrate how the distinct fluid motions 
lead to anisotropy and the spacetime metric taking the form of a Bianchi I 
solution of the Einstein equations. This follows since there is a 
spacelike privileged vector, associated with the relative flow between the 
two components in the problem. It is important to understand that, while 
this feature is natural in the multi-fluid context, it can never arise in 
the often considered multi-constituent single fluid. The multi-fluid 
hypothesis implies that each component (labelled by an index x) of the 
matter and radiation sourcing Einstein equations follows its own timelike 
vector $u_{\x}^\mu$; the relative flow between the various fluids then 
generates a privileged spacelike direction along which the Bianchi I 
solution aligns. However, it is important to recognize that it is the fluxes 
$n^\mu_\x = n_\x u^\mu_\x$, where $n_\x$ is the particle number density, 
that are the fundamental sources. In particular, a fluid can be moving 
quickly with respect to another, yet if its density is much smaller its flux can 
be negligible.

Such a choice is by no means new \cite{Sandin09} and recent work in given 
circumstances have shown, here again, the possibility of isotropisation 
\cite{HL11}, although behavior very different from the standard cosmological 
one can also be found \cite{CH10}. (A useful review on anisotropic solutions 
and their cosmological use is Ref.~\cite{TCM08}.) For instance, it has been 
suggested \cite{BT07,ABDR11,ACM11} that since Bianchi universes, seen as 
averaged inhomogeneous and anisotropic spacetimes, can have effective
strong energy condition violating stress-energy tensors, they could
be part of a backreaction driven acceleration model. 

Yet another reason for studying such cosmological models stem,
curiously, from the observations! Large angle anomalies in the Cosmic 
Microwave Background (CMB) indeed have been observed and discussed for 
quite some time \cite{SSHC04,CHSS10,Perivolaropoulos11,MEC11} and related with underlying Bianchi models \cite{PC07,Pontzen09}. It is not our
aim here to decide whether or not the data do indeed imply
some amount of anisotropy, but we shall at least assume that they
do not rule out the possibility altogether. Note in that respect
that further, currently ongoing observations of different backgrounds
will determine, for instance, if the CMB dipole is fully originating from
mere local Earth motion (and should thus be removed altogether
from the data) or if part of it is cosmological \cite{FK11}.

In order to remain close to the observationally verifiable model, we
shall concentrate on the example of the radiation to matter transition for
which, in principle, the underlying microphysics ought to be well-known,
up to the a priori necessarily negligible Dark-Matter to radiation coupling.
We then ask whether it is possible to have a cosmological epoch where 
there is a relative flow of radiation with respect to the matter, but out of 
which the expansion becomes isotropic and the relative flow dissipates. We 
will demonstrate that the short answer to this question is yes, as flux 
domination of one fluid over the other leads to an effectively one-fluid 
situation, thus yielding an effective 
Friedman-Lema\^{\i}tre-Robertson-Walker (FLRW) Universe. In essence, the 
cosmological principle appears to be satisfied on both sides of the transition, 
but the transition itself puts forward a Bianchi I behavior with a spacelike 
privileged direction. Our goal here is to, first of all, establish this possibility 
and then consider the compatibility of such a model with current observational 
data \cite{Komatsu:2010fb,Percival+10}.

On the technical side, the two-fluid nature of the problem introduces several 
terms that are not present in the one-fluid case. We will ``skew'' the 
discussion somewhat by introducing variables that were found useful in the 
stability analysis of two-fluid systems by Samuelsson \etal \cite{SLAC10}. In 
particular, we will take into account the fact that two-fluid systems have two 
speeds of ``sound'', and use causality to constrain parameter values that enter 
through the equation of state. We will also introduce the so-called cross-
constituent coupling, which occurs when the equation of state has terms 
containing both fluid densities. It is an equilibrium property and thus is non-
dissipative. While the coupling is not the main focus here, it is important for a 
follow-on analysis \cite{cpaprl} where we consider so-called two-stream 
instability. This can occur when there is a relative flow between two fluids with 
cross-constituent coupling. If a disturbance is developed on top of the relative 
flow, and the coupling is strong enough, it can become unstable if it appears 
to move, say, to the right with respect to one fluid, but to the left with respect 
to the other. In this sense, the work here has the additional purpose of 
building the ``background'', relative-flow configurations.

The outline of this paper is as follows: In Sec.~\ref{symm} we construct 
cosmologies having two Killing symmetries, with the subsequent Einstein 
tensor components presented in Sec.~\ref{etens}. Sec.~\ref{flreview} contains a 
brief review of the two-fluid formalism and how it applies in the current 
context. We also show how our formalism can be immediately employed to 
describe relativistic condensates (which reduces to the standard descriptions 
of terrestrial systems, such as superfluid helium four).  In the following
Sec.~\ref{dstrad} we show how an ideal gas in the presence of a radiation field 
leads to a system with cross-constituent coupling, and then construct a 
simpler model containing similar characteristics. Sec.~\ref{zindepen} restricts 
the analysis by removing the spatial-dependence in the metric and matter. 
(The more general set of equations are required for the two-stream instability 
analysis of \cite{cpaprl}.) This same section includes a numerical analysis 
subsection \ref{subsec:num} and ends with a discussion of the results. We 
finish with some concluding remarks in Sec.~\ref{conclude} and an appendix 
containing more details on how the equations are obtained.

\section{Cosmologies with Two Spacelike Killing Vectors} 
\label{symm}

We will choose the simplest possible two-fluid model: the relative
matter flow is in one direction (to be taken along the $z$ ``axis''),
and orthogonal to it will be two, mutually orthogonal spacelike
Killing vector fields (one along the $x$ ``axis'' and another along
the $y$ ``axis''). We will use as our $x$ and $y$ coordinates the two
parameters that naturally generate the Killing vector fields $\LX$ and
$\LY$. With this choice we have 
\bea 
    \LX = (0,1,0,0) \quad , \quad \LY = (0,0,1,0) .  
\eea 
It is also the case that 
\bea 
    0 = g_{\mu \nu} \LX {\cal Y}^\nu = g_{1 2} .  
\eea 
Finally, if we let $t$ denote the time coordinate then the two symmetries 
imply the remaining metric components are functions of only $z$ and $t$.

There is some remaining freedom in the choice of coordinate system,
\ie it can be shown that the so-called synchronous gauge ($g_{0 0} =
- 1$ and $g_{0 i} = 0$)  that reduces the metric to
\beq
    \dd  s^2 = - \dd  t^2 + g_{x x} \dd  x^2 + g_{y y} \dd  y^2 +
                  g_{z z} \dd  z^2 + 2 g_{x z} \dd  x \dd  z +
                  2 g_{y z} \dd  y \dd  z ,
\eeq
can be utilized.
Within this gauge choice there is another change of coordinates that
can be made, namely $\bar{t} = t$, $\bar{x} = \bar{x}(x,z)$,
$\bar{y} = \bar{y}(y,z)$, and $\bar{z} = z$, that sets the terms $g_{1 3}$ 
and $g_{2 3}$ to zero. The final form of the metric is thus 
\bea
    \dd s^2 = - \dd t^2 + A_x^2 \dd x^2 + A_y^2 \dd y^2 + A_z^2 \dd z^2
              , \label{fin_metric} 
\eea 
where the $A_\aleph$ ($\aleph = \{x,y,z\}$) are, as yet unknown, functions 
of $t$ and $z$. When the $z$-dependence is relaxed, the spacetime described 
by (\ref{fin_metric}) is of the well-known Bianchi I type. Although we focus 
on this case later in Sec.~\ref{zindepen}, we keep the $z$-dependence here 
because a follow-on analysis \cite{cpaprl} will need the full $z$-dependent 
equations.

\subsection{The Einstein Tensor} \label{etens}

The non-trivial Einstein Tensor coefficients can be straightforwardly
computed with the known geometric quantities given in the
Appendix. Letting a dot ``$~\dot{}~$'' and a prime ``$~^\prime{}~$'' denote, 
respectively, $\partial/\partial t$ and $\partial/\partial z$, we
have
\begin{widetext}
  \bea 
  G^t{}_t &=& - \left(H_x H_y + H_x H_z + H_y H_z\right) + \frac{1}{A^2_z} 
              \left[I^\prime_x + I^\prime_y + \left(I_x +
              I_y\right)^2 - I_x I_y - I_x I_z - I_y I_z\right] , \cr 
           && \cr
  G^x{}_x &=& - \left(\dot{H}_y + \dot{H}_z\right) - \left(H^2_y + H_y
              H_z + H^2_z\right) + \frac{1}{A^2_z} \left[I^{\prime}_y +
              \left(I_y - I_z\right) I_y\right] , \cr 
           && \cr 
  G^y{}_y &=& - \left(\dot{H}_x + \dot{H}_z\right) - \left(H^2_x + H_x H_z +
              H^2_z\right) + \frac{1}{A^2_z} \left[I^{\prime}_x + \left(I_x -
              I_z\right) I_x\right] , \cr 
           && \cr 
  G^z{}_t &=& - \frac{1}{A^2_z} \left[\dot{I}_x + \dot{I}_y + \left(H_x - 
              H_z\right) I_x + \left(H_y - H_z\right) I_y\right] , \cr 
           && \cr 
  G^z{}_z &=& - \left(\dot{H}_x + \dot{H}_y\right) - \left(H^2_x + H_x H_y +
              H^2_y\right) + \frac{I_x I_y}{A^2_z} , \label{eineqn} 
\eea
\end{widetext}
where we have introduced the ``Hubble''-like functions ($\aleph = \{x,y,z\}$) 
\beq 
    H_\aleph \equiv \frac{\dot{A}_\aleph}{A_\aleph} ,
\eeq 
and the ``inhomogeneity'' functions 
\beq 
    I_\aleph \equiv \frac{A^\prime_\aleph}{A_\aleph} .  
\eeq 
We will see below that when the $z$-dependence is dropped, the two-fluid 
energy-momentum-stress components are such that $T^x{}_x = T^y{}_y$, 
implying for the Einstein tensor $G^x{}_x = G^y{}_y$.

Clearly, not all these components can be independent of each other,
for otherwise the overall problem would be ill-posed because of too
many equations. But recall that there is the Bianchi Identity
$\nabla_\nu G^\nu{}_\mu = 0$, which for the situation here yields two
independent components:
\begin{widetext}
\bea
    0 &=& \dot{G}^t{}_t + \partial_z G^z{}_t + \left(H_x + H_y + H_z\right)
          G^t{}_t + \left(I_x + I_y + I_z\right) G^z{}_t - H_x G^x{}_x -
          H_y G^y{}_y - H_z G^z{}_z , \cr
     && \cr
    0 &=& \dot{G}^t{}_z + \partial_z G^z{}_z + \left(H_x + H_y + H_z\right)
          G^t{}_z + \left(I_x + I_y\right) \left(G^z{}_z - G^x{}_x\right) .
\eea
\end{widetext}
It is important to note that the second of these vanishes identically
when there is no $z$-dependence, because then the Einstein tensor
component $G^z_t = 0$. This means that we still need three metric degrees
of freedom.

\subsection{General Relativistic Two-fluid Formalism} \label{flreview}

We will use the formalism developed by Carter 
\cite{carter89:_covar_theor_conduc} and various collaborators (see 
Andersson and Comer \cite{andersson07:_livrev} for a  review and references). 
The fundamental fluid variables consist of two conserved number density 
four-currents, to be denoted $n^\mu_\x$. Recall that $\x$ is a constituent 
index (for which there is no implied sum when repeated). 

From the currents, we can form three scalars, namely 
$\nx^2 = - g_{\mu \nu} \nx^\mu \nx^\nu$, 
$\ny^2 = - g_{\mu \nu} \ny^\mu \ny^\nu$, and $\nxy^2
= - g_{\mu \nu} \nx^\mu \ny^\nu$. A so-called ``master'' function $-
\Lambda(\nx^2,\ny^2,\nxy^2)$ (the two-fluid analog of the equation of
state) is assumed, which plays the role of Lagrangian for the system. The energy-momentum-stress tensor is 
\beq 
    T^\mu{}_\nu = \Psi \delta^\mu{}_\nu + \nx^\mu \mu^\x_\nu + 
                  \ny^\mu \mu^\y_\nu , 
\eeq
where 
\beq 
    \Psi = \Lambda - \nx^\rho \mu^\x_\rho - \ny^\rho \mu^\y_\rho
           \label{press}
\eeq 
is the generalized pressure and 
\beq 
    \mu^\x_\nu = g_{\nu \mu} \left(\Bx \nx^\mu + \Axy \ny^\mu\right)  
\eeq 
is the chemical potential covector. It is also the momentum canonically 
conjugate to the current $\nx^\mu$. 

Formally, the $\Axy$ and $\Bx$ coefficients are obtained from $\Lambda$ 
via the partial derivatives 
\beq
    \Axy = {\cal A}^{\y \x} = - \frac{\partial \Lambda}{\partial \nxy^2}
           \quad , \quad 
    \Bx = - 2 \frac{\partial \Lambda}{\partial \nx^2} . 
\eeq 
The fact that the momentum $\mu^\x_\mu$ is not simply proportional to the 
corresponding number density current $\nx^\mu$ is a result of 
entrainment (as it is known in the neutron star literature; see, for example, 
\cite{comer03:_rel_ent}): the motion of one fluid induces a momentum in the 
other fluid, and vice versa. Entrainment vanishes if the $\Axy$ coefficient 
is zero. 

Finally, the equations for each fluid consists of a conservation equation
\beq
    \nabla_{\mu} \nx^{\mu} = 0 , \label{conseq}
\eeq
and an Euler equation
\beq
    \nx^\mu \omega^\x_{\mu \nu} = 0 , \label{euler}
\eeq
where the vorticity two-form is defined by
\beq
    \omega^\x_{\mu \nu} = 2 \nabla_{[\mu} \mu^\x_{\nu]} ,
\eeq
the square brackets indicating antisymmetrization of the enclosed indices. 
It is important to understand that the condition $\nabla_\mu T^\mu{}_\nu = 
0$ is satisfied once the equations of motion are satisfied. Contrary to the 
single-fluid case, $\nabla_\mu T^\mu{}_\nu = 0$ does {\em not} yield enough 
information to completely determine the two-fluid evolution.

Note that the above way of writing each Euler equation makes manifest its
geometric meaning as an integrability condition for the corresponding 
vorticity, a point that has been much emphasized by Carter 
\cite{carter89:_covar_theor_conduc} (see also \cite{andersson07:_livrev}). It 
also immediately supplies a formalism for superfluid condensates, since 
setting $\mu^\x_\nu = \nabla_\nu \Phi_\x$ (where $\Phi_\x$ represents the 
phase of the relevant quantum wavefunction) guarantees that the fluid vorticity 
is zero. The non-relativistic limit of the fluid equations in this case recovers 
those that are well-known for, say, helium superfluids. 

The symmetries do much to simplify the fluid equations. It is easy to see
that the vanishing of the Lie derivative of $\nx^\mu$ with respect to
$\LX$ and $\LY$ requires $\nx^\mu$ to be a function only of $t$ and $z$.
We also assume that $\nx^\mu$ is orthogonal to $\LX$ and $\LY$. The unit
four-vectors take the form
\beq
    u^\nu_\x = \left(\utx,0,0,\uzx\right)
              \quad , \quad
    \utx = \sqrt{1 + \left(A_z \uzx\right)^2} .
\eeq
The entrainment parameter becomes
\beq
    \nxy^2 = \nx \ny \left(\utx \uty - A^2_z \uzx \uzy\right) , \label{nsqxy}
\eeq
while the momenta reduce to
\bea
    \mu^\x_t &=& - \left(\Bx \nx \utx + \Axy \ny \uty\right) , \cr
              && \cr
    \mu^\x_z &=& A_z^2 \left(\Bx \nx \uzx + \Axy \ny \uzy\right) .
\eea
Finally, the components of $T^\mu{}_\nu$ are
\bea
    T^t{}_t &=& \Psi + \nx \utx \mu^\x_t + \ny \uty \mu^\y_t , \cr
             && \cr
    T^x{}_x &=& T^y{}_y = \Psi , \cr
             && \cr
    T^z{}_t &=& \nx \uzx \mu^\x_t + \ny \uzy \mu^\y_t , \cr
             && \cr
    T^z{}_z &=& \Psi + \nx \uzx \mu^\x_z + \ny \uzy \mu^\y_z ,
                \label{stresseng}
\eea
where
\beq
    \Psi = \Lambda - \nx \left(\utx \mu^\x_t + \uzx \mu^\x_z\right) - \ny
           \left(\uty \mu^\y_t + \uzy \mu^\y_z\right) .
\eeq
The remaining item required to completely specify the matter is a particular
form for the master function $\Lambda$. This we will provide in
Section~\ref{dstrad}.

We see from the above that our problem has been reduced to finding solutions
for the four matter variables $\{n_\x,u^z_\x\}$ and the three metric functions
$A_\aleph$. The conservation equations (\eqref{conseq}) now take the form
\beq
    0 = \frac{\partial}{\partial t} \left(A_x A_y A_z \nx \utx\right) +
        \frac{\partial}{\partial z} \left(A_x A_y A_z \nx \uzx\right) ,
        \label{conseqred}
\eeq
while the Euler equations reduce to
\beq
    \frac{\partial \mu^\x_t}{\partial z} =
    \frac{\partial \mu^\x_z}{\partial t}
    . \label{eueqnred}
\eeq
The remaining equations are those of Einstein, constructed from
Eqs.~(\eqref{eineqn}) and (\eqref{stresseng}).

As mentioned in the introduction, we introduce some new variables that are 
convenient for the multi-fluid analysis. Since there are two fluids we have 
the well-established result of two modes of ``sound'' propagation 
\cite{carter89:_covar_theor_conduc}; namely,
\beq
    \cx^2 \equiv \frac{\partial \ln \mu^\x}{\partial \ln \nx} .
                 \label{baress}
\eeq
These are ``bare'' in the sense that they only equal the local wave speed 
when there are no interactions between the fluids \cite{SLAC10}. A measure 
of the interactions are the cross-constituent couplings defined---slightly 
modified from \cite{SLAC10}---as
\beq
    \Cxy \equiv \frac{\partial \ln \mu^\x}{\partial \ln \ny} =
    \frac{\mu^\y \ny}{\mu^\x \nx} \Cyx , \label{cccoup}
\eeq
where, if we set the entrainment to zero,
\beq
    \mu^\x \equiv - u^\nu_\x \mu^x_\nu = \Bx \nx .
\eeq
The $\Cxy$ represent a key channel through which the two fluids ``see'' 
each other (especially when the entrainment is zero) \cite{SLAC10,cpaprl}. 

Some final words on this set-up is about our frame of reference. We have 
chosen a frame that is not attached to either fluid. One might expect it 
would be easier to work in either of the fluid rest-frames, but this is actually
not the case. Starting with the metric in Eq.~(\ref{fin_metric}), we 
can show that ``jumping'' on a fluid rest-frame introduces a shift vector 
into the metric. 

Let $\bar{x}^\mu$ be the rest-frame coordinates of, say, the $\x$-fluid. 
We can assume that the coordinate transformation does not involve the 
orthogonal pair $\{x,y\}$, so that $\bar{t} = \bar{t}(t,z)$, $\bar{x} = x$, 
$\bar{y} = y$, and $\bar{z} = \bar{z}(t,z)$, which guarantees 
$\bar{u}^x_\x = \bar{u}^y_\x = 0$. What we want is $\bar{u}^z_\x = 0$, 
which implies
\beq
    \frac{\partial \bar{z}}{\partial t} = - 
    \frac{\partial \bar{z}}{\partial z} \frac{u^z_\x}{u^t_\x} ,
\eeq
and therefore $\bar{z}$ must depend on both $t$ and $z$. We can now assume 
that $\bar{t} = t$. However, the change of coordinates also 
affects the metric; in particular,
\beq
    \bar{g}^{t z} = - \frac{\partial \bar{z}}{\partial t} \neq 0 .  
\eeq

\section{A Cosmological Two-fluid Scenario: Matter and Radiation}
\label{dstrad}

When the particle species of a fluid has mass $m^\x$, it can be useful to 
separate out from $\Lambda$ mass density terms; namely,
\beq
    \Lambda = - m^\x n_\x - m^\y n_\y - {\cal E}(\nx^2,\ny^2,\nxy^2) \ ,
\eeq
where ${\cal E}$ contains other information about the fluid thermodynamics, 
and relative motion effects. The two-fluid cosmology we have in mind has a 
combination of ``matter'', with mass $m^\x = m$, and ``radiation'', which   
means $m^\y = 0$. We assume a non-zero cross-constituent coupling 
and zero entrainment. One of our conserved currents is the total particle 
flux of the matter. Since we are ignoring dissipation in the flows, we can use 
the total entropy flux of the system as our other conserved current. The bulk 
of this is due to the radiation. To simplify the notation, we set $\nx = n$, 
$\ny = s$, $\mu^\x \equiv \mu$, and $\mu^\y \equiv T$, which is the 
temperature.

To see how a cross-constituent term can come about, consider the usual way 
of combining a (non-relativistic) gas and radiation in the energy density 
and pressure:
\bea
    \rho &=& m n + \frac{3}{2} n T + \alpha T^4 , \\
          && \cr
    p &=& n T + \frac{1}{3} \alpha T^4 ,
\eea
where $\alpha$ is constant. We take as our fundamental thermodynamic 
variables $n$ and $s$, and so the temperature, obtained as $T = \partial 
\rho/\partial s$, is a function of both. Hence, the ideal gas contribution will 
generate a cross-constituent coupling (\cf Eq.~(\ref{cccoup})). Even if we 
take the temperature as fundamental, there would still be its coupling with 
$n$. 

Writing $T$ in terms of $\{n,s\}$ explicitly is not tractable. So for the purpose 
at hand, it is perhaps clearer to consider a simpler, algebraic construction 
where the dependence is explicit. With that in mind, we will use a master 
function of the form
\beq
    \Lambda = - m^* n - \kappa_\s s^{4/3} , \label{lam_mod} 
\eeq
where we have placed a polytropic coupling to the entropy in an effective
mass $m^*$ for the matter; namely,
\beq
     m^* = m + \tau_{\n \s} n^{\sigma_\n - 1} s^{\sigma_\s} ,
\eeq
where $\sigma_\n \geq 1$, $\sigma_\s \geq 1$, and $\tau_{\n \s}$ are
constants. The remaining fluid variables are
\bea
    \Psi &=& \frac{1}{3} \kappa_\s s^{4/3} + \left(\sigma_\n + \sigma_\s - 1
             \right) \left(m^* - m\right) n , \\
         && \cr
    \mu &=& m + \sigma_\n \left(m^* - m\right) , \\
         && \cr
    T &=& \frac{4}{3} \kappa_\s s^{1/3} + \sigma_\s \left(m^* - m\right)
               \frac{n}{s} , \label{temp} \\
         && \cr
    \Cns &=& \frac{\sigma_\s}{\sigma_\n - 1} \cn^2 , \label{dustrad}
\eea
where
\bea
    \mu \cn^2 &=& \sigma_\n \left(\sigma_\n - 1\right) \left(m^* - m\right)
                  , \label{nsndspd} \\
               && \cr
    T \cs^2 &=& \frac{4}{9} \kappa_\s s^{1/3} + \sigma_\s \left(\sigma_\s -
                1\right) \left(m^* - m\right) \frac{n}{s} .
                \label{ssndspd}
\eea

There are a few comments to be made about this construction. In order to 
have a model that cools as it expands, we see that $n \to 0$ and $s \to 0$ 
which also means $m^* \to m$. This also ensures that the ``dust'' limit of the 
standard cosmological scenario $\mu \to m$ and $c_\n \to 0$ is achieved. 
Finally, we recover the usual result for the massless fluid of $s \propto T^3$ 
and $\cs^2 \to 1/3$. In fact, if we eliminate the second term in 
(\eqref{ssndspd}) using (\eqref{temp}), we find 
\beq 
    s = \sigma T^3 
          \quad \hbox{and} \quad 
   \sigma = \left[\frac{3 \left(\sigma_\s - 1 - \cs^2\right)}
                  {4 \left(\sigma_\s - 4/3\right) \kappa_\s}\right]^3 .
                 \label{st3}
\eeq 
It is also worthwhile to consider the other direction of the evolution, which is 
that back to the past, where the universe contracts and heats up to the point 
where the temperature scale is much higher than that of the mass scale.

\section{Homogeneous Background} \label{zindepen}

Assuming that the background is only time-dependent, then the two
matter equations (\eqref{conseqred}) and (\eqref{eueqnred}) imply (for
$\x = \{\n,\s\}$) 
\beq 
    A_z \mu^\x \vx = \Mx 
        \quad , \quad 
    A_x A_y A_z \nx \sqrt{1 + \vx^2} = \Nx ,
          \label{matsln}
\eeq
where $\Mx$ and $\Nx$ are constants and we have introduced $\vx = A_z
u^z_\x$. One can also show that $T^z{}_t = 0$ is automatically guaranteed
by (\eqref{matsln}), provided that the integration constants satisfy
\beq
    \Mn \Nn + \Ms \Ns = 0 . \label{int_consts_rel}
\eeq

Equation (\eqref{matsln}) allows, in principle, to write $\nx$ and $\vx$ in
terms of $A_\aleph$, which can be put into the Einstein equations to get a
closed system of equations. But, since we will be solving the equations
numerically, it is actually easier to use the original differential equations, 
which can be shown to take the form
\begin{widetext}
\beq
    \left(\begin{array}{cc}
    1 - \displaystyle\frac{\cn^{2} \vn^2}{1 + \vn^2} &
    - \displaystyle\frac{\Cns \vn^2}{1 + \vn^2} \\ \\
    - \displaystyle\frac{\Csn \vs^2}{1 + \vs^2} &
    1 - \displaystyle\frac{\cs^{2} \vs^2}{1 + \vs^2}
    \end{array}\right)
    \left(\begin{array}{c}
    \displaystyle\frac{\dot{n}}{n} \\ \\
   \displaystyle \frac{\dot{s}}{s}
    \end{array}\right) =
    - \left(\begin{array}{c} H_x + H_y + \displaystyle\frac{H_z}{1 + \vn^2} 
    \\ \\
    H_x + H_y + \displaystyle\frac{H_z}{1 + \vs^2}
    \end{array}\right)
\label{densities}
\eeq
for the densities and
\beq
    \left(\begin{array}{c}
    \displaystyle\frac{\dot{V}_\n}{\vn} \\ \\
    \displaystyle\frac{\dot{V}_\s}{\vs}
    \end{array}\right) =
    - \left(\begin{array}{cc}
    \cn^2 & \Cns \\ \\
    \Csn & \cs^2
    \end{array}\right)
    \left(\begin{array}{c} \displaystyle\frac{\dot{n}}{n} \\ \\
    \displaystyle\frac{\dot{s}}{s} \end{array}\right) -
    \left(\begin{array}{c} H_z \\ \\ H_z \end{array}\right)
\label{velocities}
\eeq
for the velocities.

To solve for the metric we use the definition of $H_\aleph$ and three of the
Einstein equations (setting $\GN= \Mp^{-2}$ with $\Mp$ the Planck mass)
to evolve $\{A_\aleph,H_\aleph\}$ as follows:
\bea
    \dot{H}_x &=& - H^2_x + H_y H_z - \frac{4 \pi}{\Mp^2} \left[\mu n 
                  \left(1 + 2 \vn^2\right) + T s \left(1 + 2 \vs^2\right)
                  \right] , \label{Hx}\\
               && \cr
    \dot{H}_y &=& - H^2_y + H_x H_z - \frac{4 \pi}{\Mp^2}  \left[\mu n 
                  \left(1 + 2 \vn^2\right) + T s \left(1 + 2 \vs^2\right)
                  \right] , \label{Hy} \\
               && \cr
    \dot{H}_z &=& - H^2_z + H_x H_y - \frac{4 \pi}{\Mp^2} 
                  \left(\mu n + T s\right) , \label{Hz} \\
               && \cr
    \dot{A}_\aleph &=& H_\aleph A_\aleph . \nonumber
\eea
The so-called Hamiltonian constraint $G^t{}_t = \displaystyle
\frac{8 \pi}{\Mp^2}  T^t{}_t$ is
\beq
    H_x H_y + H_x H_z + H_y H_z = \frac{8 \pi}{\Mp^2}\left(- \Lambda + 
              \mu n \vn^2 + T s\vs^2\right) . \label{Hcons}
\eeq
\end{widetext}

The initial conditions therefore consist of four matter, and six metric 
initial conditions; \ie the set 
$\{n(t_0),s(t_0),\vn(t_0),\vs(t_0),A_\aleph(t_0),H_\aleph(t_0)\}$,
where $t_0$ is the initial time.

\begin{figure}[h]
\centering
\includegraphics[width=8.0cm,clip]{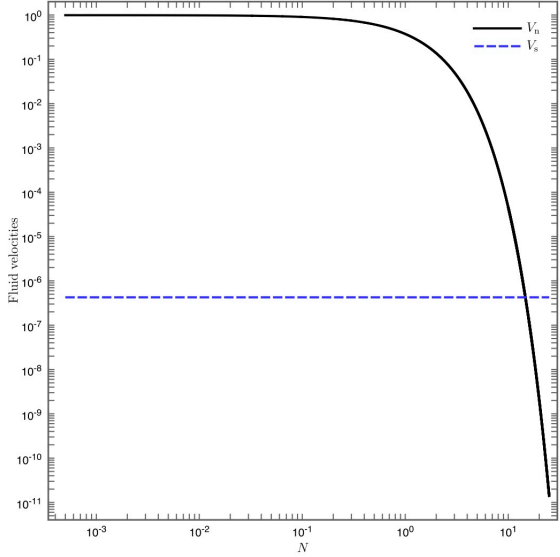}
\caption{Velocities derived from Eq.~(\ref{velocities}), as functions
of the e-fold number $N$ (defined in the main text). The underlying parameter 
values for this plot are $\kappa_\s = 1$, $\tilde\tau = 0.1$, 
$\sigma_\n = 1.1$, and $\sigma_\s = 1.1$. The initial values are such
that $n(0) = 3.9\times 10^{-7}$, $s(0) = 1$, $V_\n (0)= 0.99$, $V_\s (0)
= -4.25\times 10^{-7}$, $A_\aleph(0) =2$, and $H_\aleph(0) = 2.89$ (for each 
$\aleph$). This figure illustrates that $V_\n$, although initially very 
large, rapidly decays to zero while $V_\s$ remains essentially negligible, 
and almost constant, at all times.}
\label{fig:vel}
\end{figure}

\subsection{Preliminaries: The Matter Quadratures}

It is useful at this point to apply the results of Eq.~(\eqref{matsln}). For the 
matter we can write
\bea
    n &=& \frac{1}{A_x A_y A_z} \frac{\Nn}{\sqrt{1 + \Mn^2/\left(A_z \mu
          \right)^2}} , \label{nexact}\\
       && \cr
    \vn &=& \frac{\Mn}{A_z \mu} .
\eea
The first indicates that the metric coefficients must grow with time if both
$n \to 0$ and $\mu \to m$ (the conditions for cooling); that is, we can be
sure that our model allows for both expansion and cooling. The second 
relation therefore shows that $\vn \to 0$ with time.

A similar analysis  for the entropy fluid is complicated by the fact that it is 
massless, and thus the associated chemical potential (\ie the temperature) 
can go to zero. In particular, it is not clear a priori that the entropy fluid 
velocity
\beq
    \vs = \frac{\Ms}{A_z T}
\eeq
remains finite, \ie whether or not $A_z T$ grows with time. Actually, the
entropy relation
\beq
    s = \frac{1}{A_x A_y A_z} \frac{\Ns}{\sqrt{1 + \Ms^2/
        \left(A_z T\right)^2}} \label{entsln}
\eeq
shows that $s \to 0$ even if $A_z T \to 0$. In fact, we see that
\beq
    0 \leq A_x A_y A_z s \leq \Ns .
\eeq
As for the behavior of $A_z T$, we can show that substituting 
Eq.~(\eqref{entsln}) into Eq.~(\eqref{st3}) results in a cubic for 
$\left(A_z T/\Ms\right)^2$, which is
\beq
    \left(\frac{A_z T}{\Ms}\right)^6 + \left(\frac{A_z T}{\Ms}\right)^4 - 
     \left(\frac{A_z^2}{A_x A_y}\right)^2 
           \left(\frac{\Ns}{\sigma \Ms^3}\right)^2 = 0 .
\eeq
If the FRLW solution is obtained in the late time limit, then the last 
term tends to a constant, and hence $A_z T$ as well. This shows that 
$\vs \to const$ and $u^z_\s \to 0$. In fact, we see in Fig.~\ref{fig:vel} 
that this is precisely the case. The bottom line is that this form of 
model is such that the expansion can become isotropic, and damp out the 
three-velocities of each fluid.

\subsection{Numerical Results}
\label{subsec:num}

Numerically solving the system of Eqs.~(\ref{densities})--(\ref{Hz}) requires 
that we rewrite those in terms of dimensionless quantities. Rescaling the time 
variable to $t\to m^2 t/\Mp$, and denoting by an overdot the derivative with 
respect to this new dimensionless time, we set 
\beq
x \equiv \frac{n}{m^3}, \ \ \ \ \ y\equiv \frac{s}{m^3}\ , \ \ \ \ \ \tilde\mu 
         \equiv 
      \frac{\mu}{m} \ \ \ \ \hbox{and} \ \ \ \ \tilde T\equiv \frac{T}{m} ,
\label{vardim}
\eeq
together with 
\beq
    \tilde\tau \equiv \tau_{\n\s} m^{3(\sigma_\n -\sigma_\s)-4} ,
\eeq
yielding
\bea
   \tilde\mu &=& 1 + \sigma_\n \tilde\tau x^{\sigma_\n-1}y^\sigma_\s ,\\
   \tilde T &=&\frac43 \kappa_\s y^{1/3} + \sigma_\s \tilde\tau x^\sigma_\n 
   y^{\sigma_\s-1} ,
\eea
showing that the system is fully determined provided the two
arbitrary dimensionless constants $\kappa_\s$ and $\tilde\tau$
are given.

To make comparison with standard cosmology clearer, we
further rewrite the Bianchi I metric Eq.~(\ref{fin_metric})
in the form
\beq
    \dd s^2 = -\dd t^2 + a^2(t)\left(\ex^{2\beta_x} \dd x^2 +
              \ex^{2\beta_y} \dd y^2+\ex^{2\beta_z} \dd z^2\right),
              \label{BI}
\eeq
with $\displaystyle \sum_\aleph \beta_\aleph = 0$, thus defining the scale 
factor $a(t)$. The relations to pass from Eq.~(\ref{fin_metric}) to 
Eq.~(\ref{BI}) are then
\beq
    a^3 = A_x A_y A_z \ \ \ \ \hbox{and} \ \ \ \ \ \beta_x =
          \frac13 \ln \frac{A_x^2}{A_y A_z} ,
\eeq
with similar relations for $\beta_y$ and $\beta_z$ obtained by
circular permutations of the indices $(x,y,z)$. The so-called shear 
variables \cite{PPJPU} are given by
\beq
   \sigma_\aleph \equiv \dot\beta_\aleph \ex^{2\beta_\aleph} 
                 = \frac{A_\aleph^2}{a^2} \dot\beta_\aleph . \label{shear}
\eeq

We can rewrite the equations of motion in terms of the e-fold number $N$,
defined through 
\beq
    a(t) = \ex^N,
\eeq
by using the relation
\beq
    \frac{\dd}{\dd t} = \frac13 \left( H_x+H_y+H_z\right)\frac{\dd}{\dd N}.
\eeq
The figures that illustrate our results all use this parameter $N$ for the 
horizontal axis.

\begin{figure}[ht]
\centering
\includegraphics[width=8cm,clip]{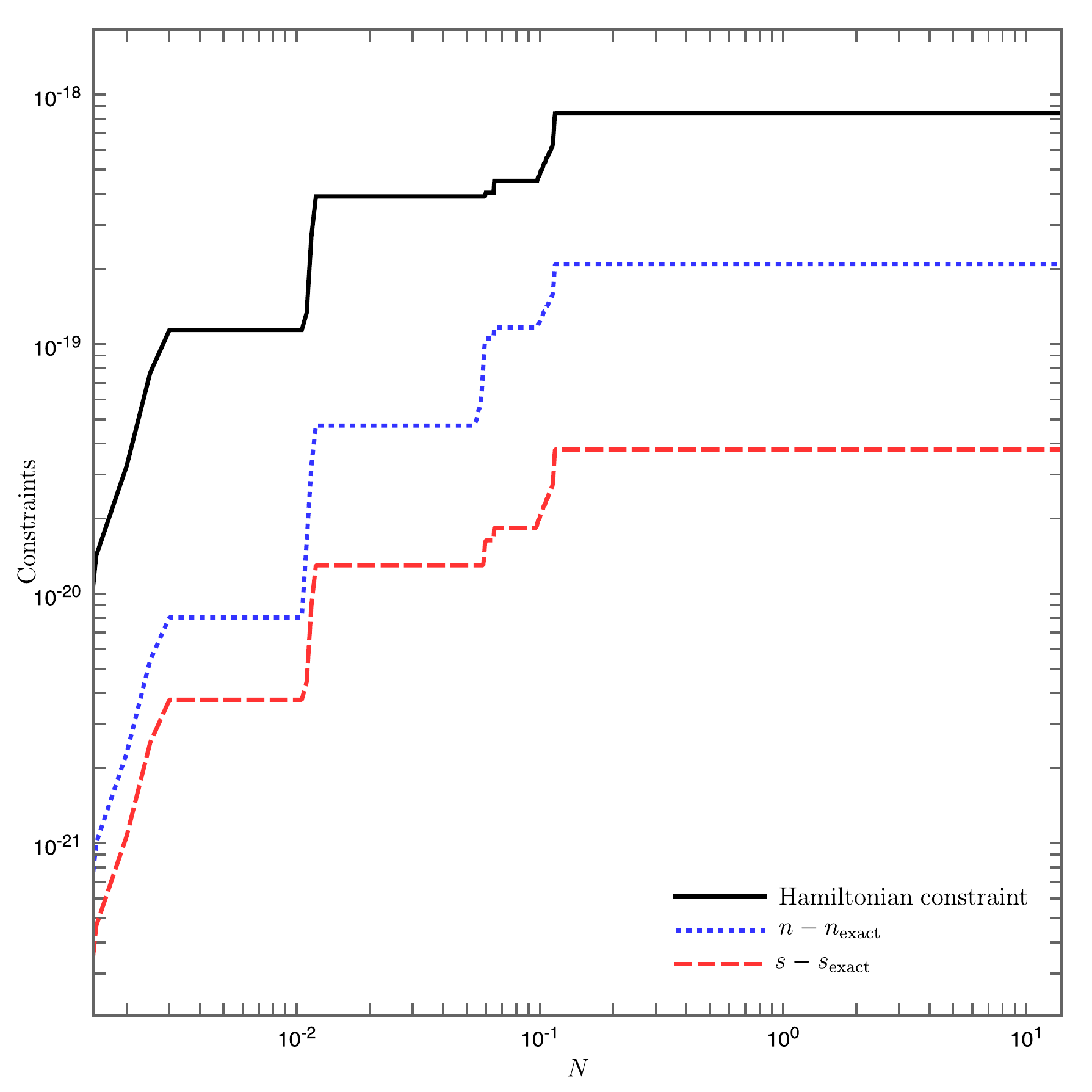}
\caption{Evaluation of the overall numerical errors associated with the 
solution of 
Eqs.~(\ref{densities})--(\ref{Hz}): with the same parameters as in the previous 
figures, we show here the levels at which Eqs.~(\ref{nexact}), (\ref{entsln}) and 
(\ref{Hcons}) are satisfied. The most error-prone situation, at the end of the 
calculation, still satisfies the constraints to better than $10^{-18}$.}
\label{fig:err}
\end{figure}

A realistic model having two FLRW phases connected by a Bianchi I
transition is realized through numerical solutions of 
Eqs.~(\ref{densities})--(\ref{Hz}). We use the exact solutions of Eqs.~(\ref{nexact}) 
and (\eqref{matsln}), together with the Hamiltonian constraint (\ref{Hcons}) as a 
measure of the numerical error. This is given in Fig.~\ref{fig:err}, which shows the 
relative error, for our particular choice of parameters, to be limited to at most 
$10^{-18}$.

\begin{figure}[ht]
\centering
\includegraphics[width=8cm,clip]{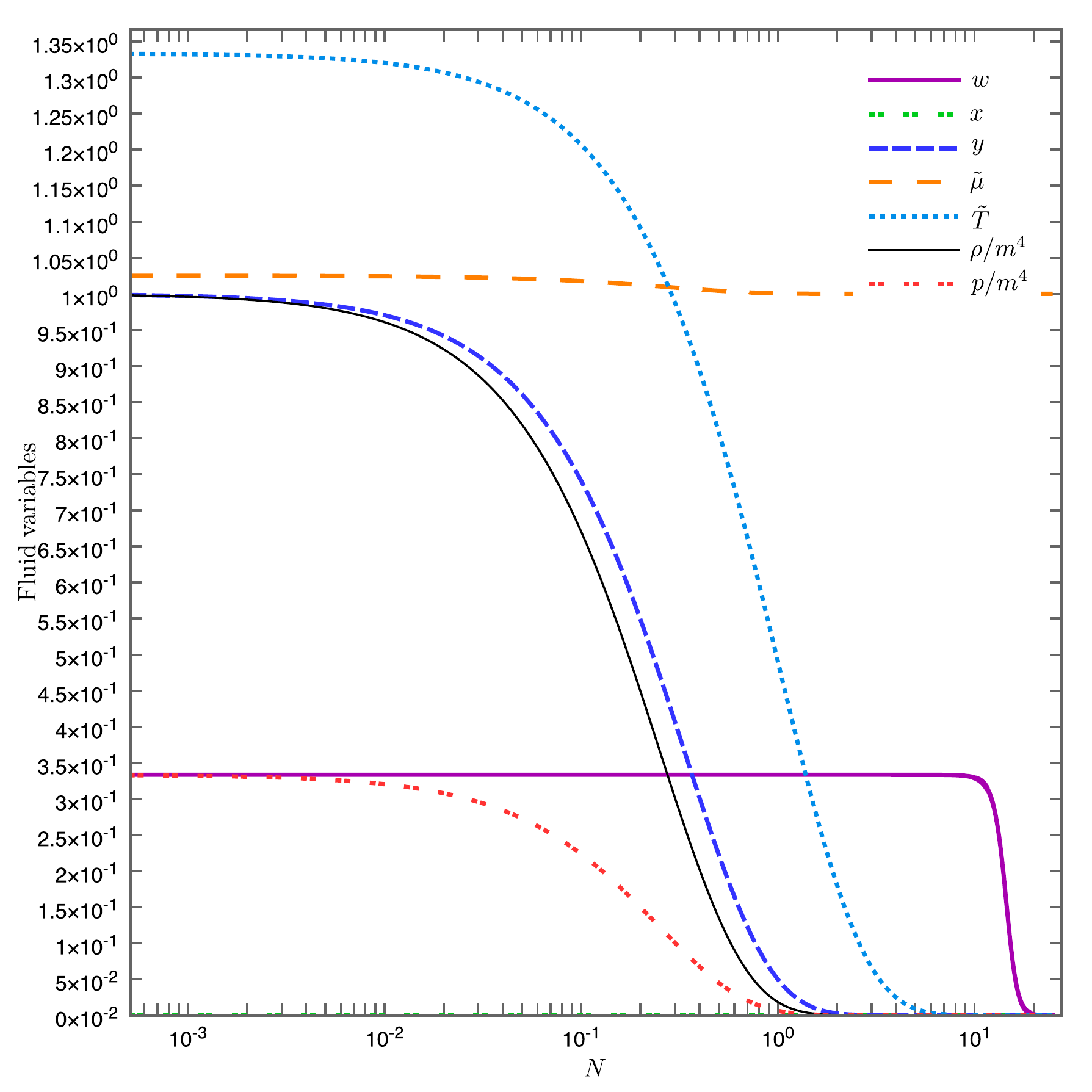}
\caption{Background fields for the same parameters as Fig. \ref{fig:vel}. \
The thick solid line corresponds to the state parameter $w$, \ie the ratio of
$\rho$ (thin full line) and $p=\Psi$ (thick double-dotted line), the number
density $x\propto n$ is essentially negligible at these scales, while the 
density $\rho$ is dominated by the contribution of the entropy $y\propto s$ 
(thick dashed line); the conjugate variables, namely the rescaled 
temperature $\tilde T$ (thick dotted line) and matter chemical potential 
$\tilde\mu$ (thick long dashed line) both decrease, with $\tilde\mu\to 1$ as 
expected.}
\label{fig:var}
\end{figure}

Fig.~\ref{fig:var} shows the behavior of the fluid variables with $N$ around 
the radiation to matter transition, \ie with the state parameter $w$ smoothly 
varying from its initial value of $\frac13$ to zero. The rescaled number 
density $x$ is found to be negligible throughout, even though its 
contribution to the energy density eventually dominates. The rescaled 
entropy $y$ provides, roughly, all of the energy density $\rho$ initially and 
for most of the transition, but eventually becomes negligible, as 
expected. Finally, the temperature $\tilde T$ is seen to decay to zero, 
while the rescaled chemical potential $\tilde \mu$ asymptotically takes its 
fiducial value unity.

\begin{figure}[ht]
\centering
\includegraphics[width=8cm,clip]{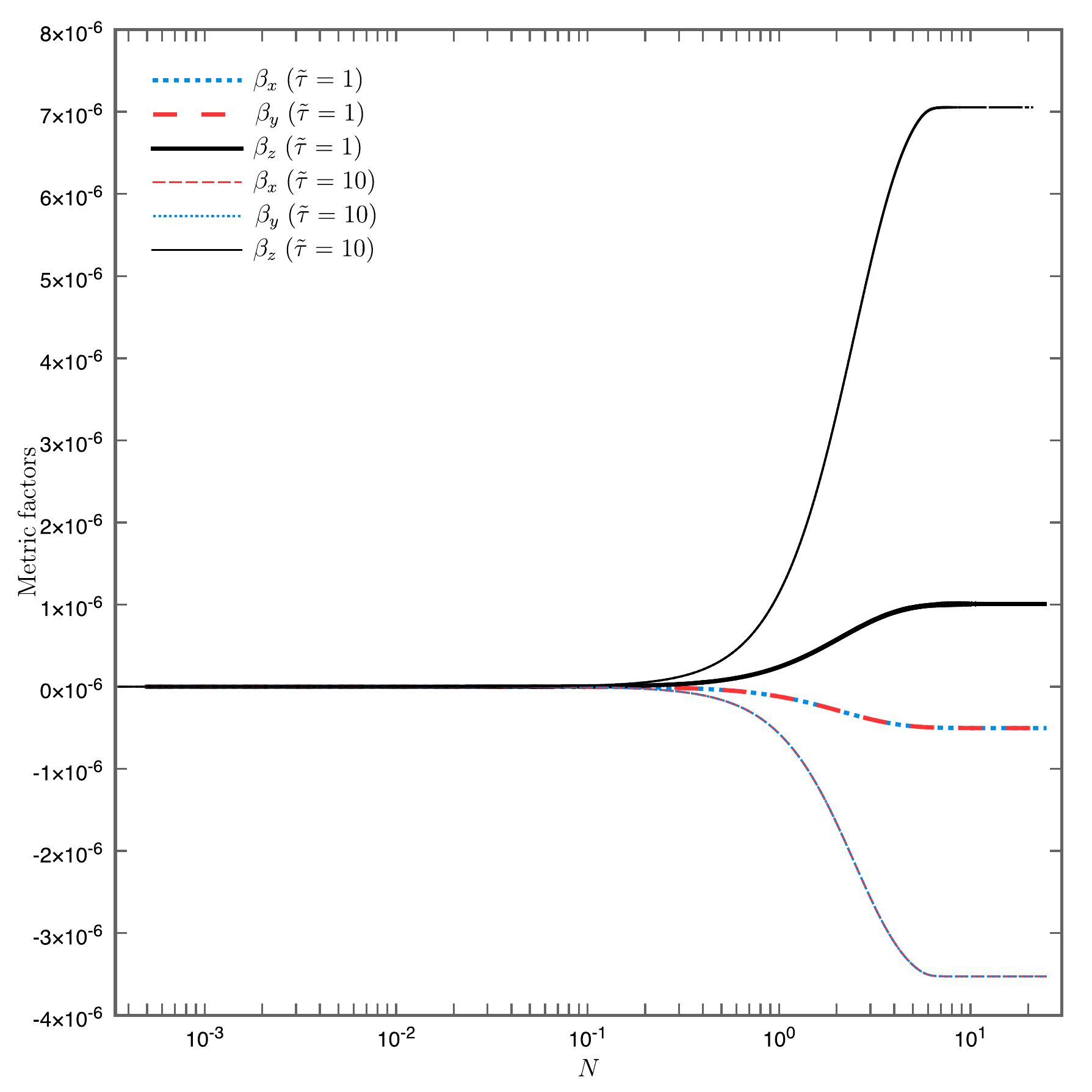}
\caption{Metric coefficients $\beta_i$ as functions of the e-fold number
$N$, for the same parameters as Fig.~\ref{fig:vel}, except that 
$\tilde{\tau} = 1,10$. The full thick line represents $\beta_z$, while the other two 
(dotted and dashed lines) stand for $\beta_x$ and $\beta_y$ respectively, 
satisfying $\beta_x=\beta_y=-\frac12 \beta_z$, in agreement with our setting.}
\label{fig:betas}
\end{figure}

\begin{figure}[ht]
\centering
\includegraphics[width=8cm,clip]{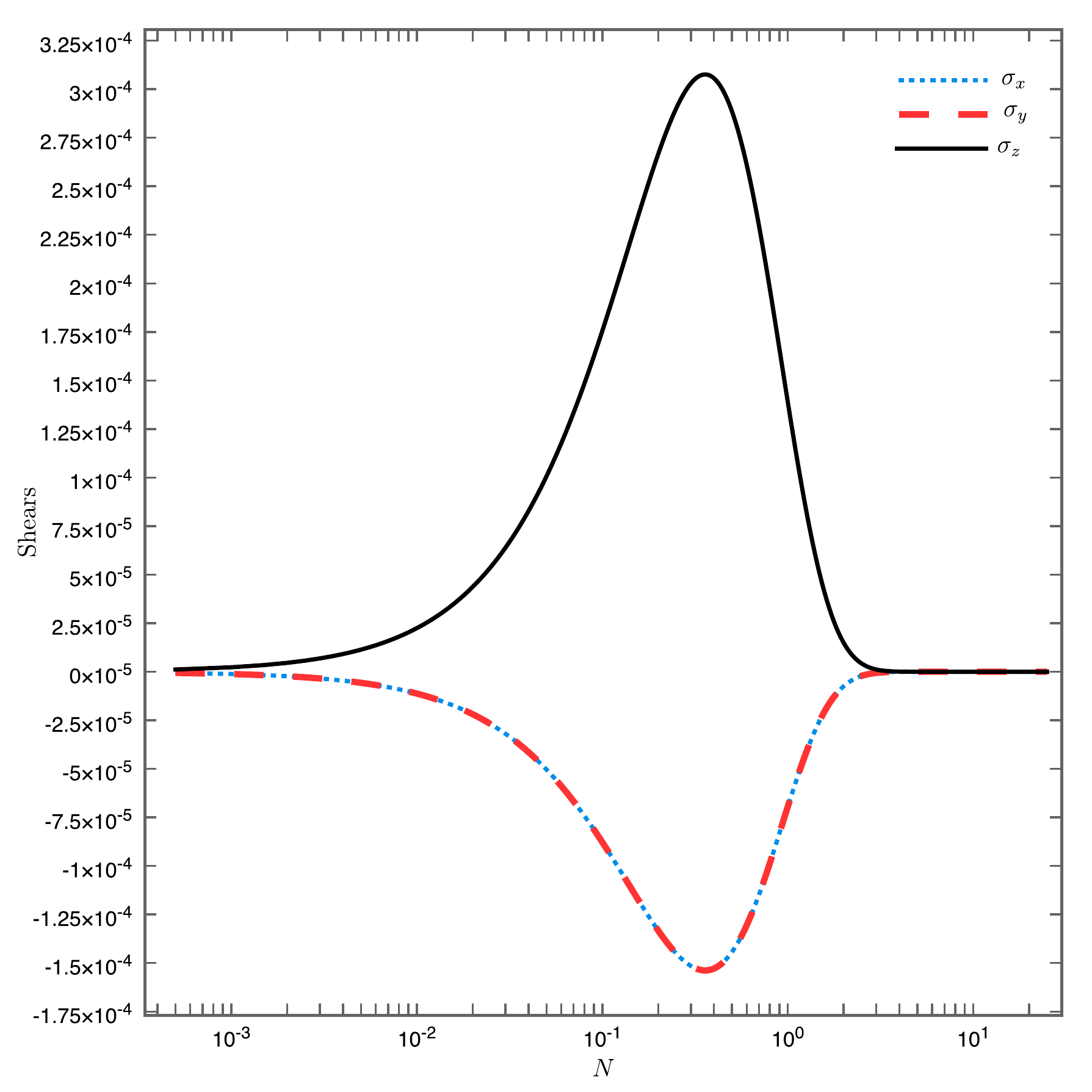}
\caption{Shear functions $\sigma_i$ as functions of the e-fold number
$N$, for the same parameters as Fig. \ref{fig:vel}, except that 
$\tilde{\tau} = 1,10$. The full thick line represents $\sigma_z$, while the other 
two (dotted and dashed lines) stand for $\sigma_x$ and $\sigma_y$ 
respectively. Because of the relation between the metric coefficients $\beta_i$, it 
turns out that the shears satisfy a similar relation, namely 
$\sigma_x=\sigma_y=-\frac12 \sigma_z$. We also see that the anistropies 
increase as the coupling $\tilde{\tau}$ is increased.}
\label{fig:sigmas}
\end{figure}

The behavior of the metric and the shears are displayed in 
Figs.~\ref{fig:betas} and \ref{fig:sigmas}, respectively. The beta 
coefficients change from being initially equal to each other (zero in the 
numerical calculation), to final constant values. With a rescaling of the 
spatial coordinates we can absorb these constants so as to return to the usual 
FLRW metric. Here we use the same parameters as before, except that we 
have taken $\tilde{\tau} = 1,10$. The reason is illustrated in Fig.~\ref{fig:sigmas}, 
which shows that the shears $\sigma_\aleph$, initially vanishing (because we 
start with a FLRW radiation dominated phase), increase first during the 
transition, reach a maximum and eventually decrease to vanishingly small 
values, which is expected for the final FLRW matter dominated epoch. As one 
might expect, as the coupling $\tilde{\tau}$ is increased, the anisotropies 
increase.

\begin{figure}[ht]
\centering
\includegraphics[width=8cm,clip]{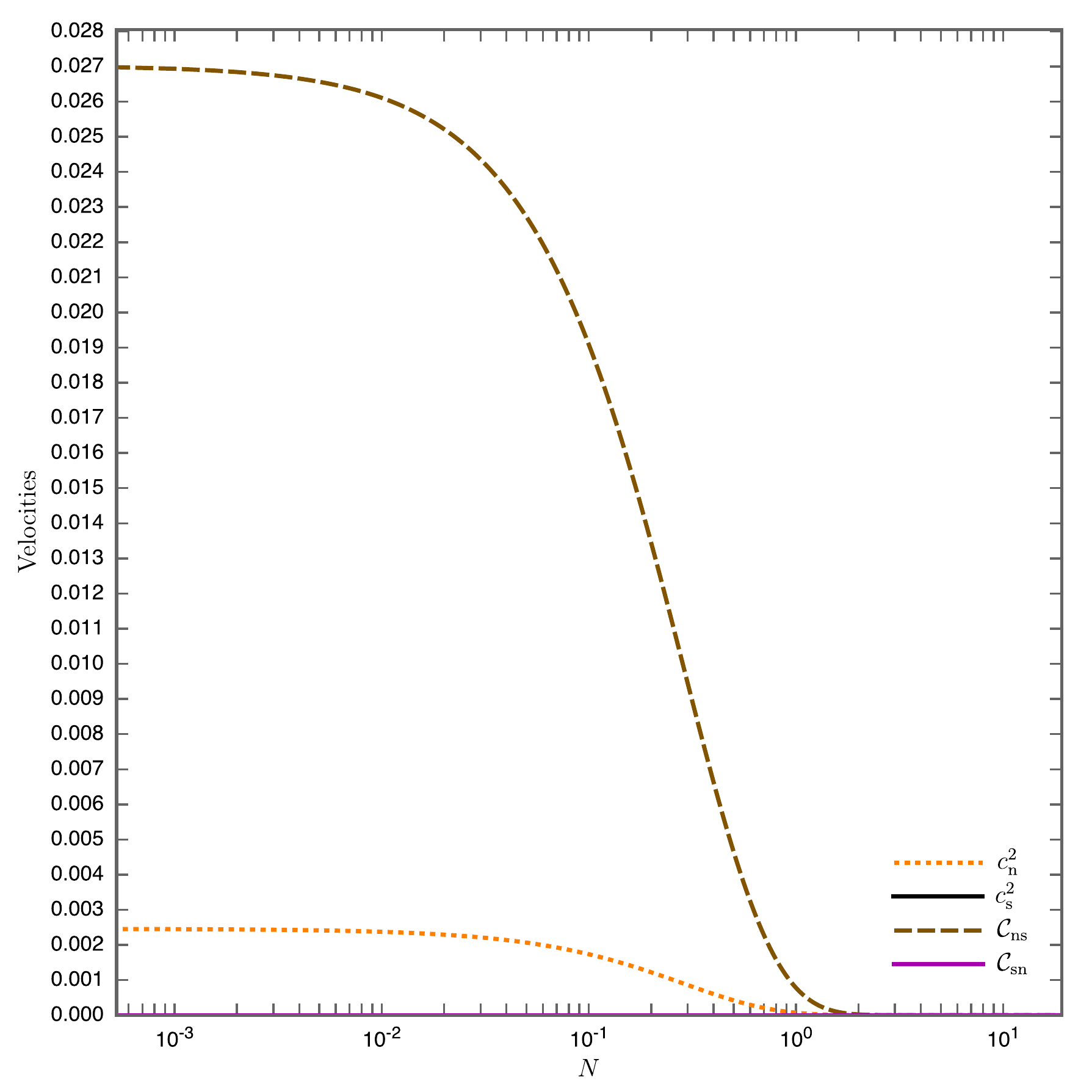}
\caption{The bare sound speeds $c^2_\n$, $c^2_\s$, and cross-constituent 
couplings $\mathcal{C}_{\n\s}$ and $\mathcal{C}_{\s\n}$ as functions of the 
e-fold number $N$, for the same parameters as Fig.~\ref{fig:vel}. Since 
$c^2_\s = 1/3$ throughout, it is not shown in the figure. Both $c^2_\n$ 
(dotted line) and $\mathcal{C}_{\n\s}$ (dashed line) are smoothly decaying 
functions of time, while $\mathcal{C}_{\s\n}$ (full line, hardly visible on 
the figure) is essentially negligible at all times.}
\label{fig:iv}
\end{figure}

Finally, Fig.~\ref{fig:iv} shows the time evolution of the sound speeds
$c^2_\n$ and $c^2_\s$, as well as $\mathcal{C}_{\n\s}$ and 
$\mathcal{C}_{\s\n}$. We find that $c^2_\s = 1/3$ throughout, although 
$c^2_\n$ is modified in the radiation era. Both cross-coupling terms decrease 
with time. This, along with the vanishing of the relative flow, is an important 
result for a two-stream instability analysis, for it implies that the conditions for 
instability are naturally eliminated by the overall expansion of the universe.

With the solution at hand, it is now possible to return to the
original equations and understand what is taking place during
the transition. Originally, we set initial conditions in the radiation
era, for which $Ts\gg \mu n$, with the extra requirement that
$V_\s\ll 1$: this means that the evolution of the three Hubble
functions, and hence of the scale factors, will be identical, so the 
shears vanish and we are in a FLRW phase. Then, as the product
$\mu n$ begins to grow, with $Ts$ decreasing, the matter velocity,
provided it was large enough to begin with (and we see numerically
that we need to set it very close to unity in order to have a visible
effect) is still large enough that the corresponding term becomes
important and the scale factors begin to evolve in different ways.
Finally, even this velocity becomes sufficiently small with respect
to unity that one recovers the FLRW symmetry expected for the
matter dominated epoch. As to why  the matter velocity can be large, 
and yet the universe be radiation dominated, is because it is the flux 
that enters the Einstein equations; a large velocity can be compensated 
for by a small number density.

\section{Closing Remarks}
\label{conclude}

A main goal of this work was to develop for cosmology the  general 
relativistic, multi-fluid model (derived from a variational 
formalism) that has so far been used mostly for neutron star astrophysics. 
While we considered only a two-fluid system, the formalism itself can, 
in principle, handle a number of different fluids. As it comes from 
an action, coupling to other fields can be imposed in more or less 
standard ways. For example, electromagnetism can be incorporated via 
the usual gauge coupling, thus allowing for plasmas and their effects 
on the system. We also demonstrated how relativistic condensates 
follow automatically because the formalism is written in terms of 
the conjugate momenta, and simply setting the momenta to be gradients 
of scalars automatically insures zero vorticity.

The two-fluid model we introduced is valid for applications
in cosmology. Even though the relative motions were anti-aligned, 
the model illustrated behavior that one might expect from a close 
examination of the radiation-to-matter transition. The main 
task was to build a model in which both the radiation and the matter 
dominated eras were describable by means of an FLRW metric, as 
necessary to fit nucleosynthesis, CMB, and large scale structure 
formation data \cite{Komatsu:2010fb,Percival+10}. We found that  
such a situation could easily be implemented, provided the relative 
fluid velocity is large enough at the transition time, an assumption 
that needs to be justified on the basis of primordial cosmology models. 
Indeed, in the now well-established framework of inflation 
\cite{LNP2008}, it is tremendously difficult for the Universe to 
remain with any relevant amount of left-over anisotropy: in fact, 
inflation was precisely invented to, among its expected outsets, 
remove any primordial anisotropy!

It should be recalled at this point that the model presented above 
is the simplest set-up of what might be envisioned for the transition 
itself, for we have not taken into account, for example, the 
non-conservation of the photon number through its coupling with 
luminous matter, matter flows with more than one constituent, or 
relative flows at arbitrary angles. Obviously, one would not be 
too surprised if comparison to observational data indicated the 
need for a more elaborate model. Note also that we have assumed 
here the matter fluid to have only one flux-component. This may not be 
a reasonable assumption; it is, however, largely a 
scale-dependent statement.

There are not so many ways to produce such primordial
anisotropy. Among the most natural  are perhaps models
based on some amount of non trivial electromagnetic phenomena
taking place during very early epochs. Consistent with large
scale astrophysical observations of $\gamma-$ray halos 
around active galactic nuclei \cite{AK10}, the existence of
relatively intense intergalactic magnetic fields of the order
of $10^{-15}$~G have been deduced, whose formation is
expected to be of primordial origin. Some inflationary models
\cite{AS06} are able to produce such large scale magnetic fields,
that are statistically isotropic. It requires a special effort to
construct a so-called ``hairy" universe \cite{WKS09} in
which the resulting magnetic field (or any other gauge field 
coherent over large distances) points to a privileged spatial
direction; off-diagonal $TB$ and $EB$ spectra could be
induced by such models \cite{WKS11}, hence providing an
observational means to validate them.

A special spatial direction can also exist in more radical
scenarios. In one such model, for instance, a planar domain wall
remains all through the inflation phase, thereby breaking the rotational
invariance of the final perturbation power spectrum \cite{WWH11}. 
Multifield inflation can also serve that purpose by producing 
vorticity, although at second order in the relevant perturbations \cite{CMM09}.

Another way to induce a non FLRW universe (perhaps the simplest)
is to start with a theory having a built-in privileged timelike
vector with which the dominant fluid may not necessarily align; examples
are provided by the Ho\v{r}ava--Lifshitz setup \cite{HL2009,BPS10},
originally aimed at renormalizing gravity, and the Einstein-\ae{}ther
theory \cite{Jacobson:2008aj}.\footnote{In the IR limit, these theories turn 
out to be equivalent \cite{Jacobson2010}.} Depending on the initial
conditions, their solution can actually also relax to the usual FLRW
solution, the fluid unit vector then aligning dynamically with the
\ae{}ther vector \cite{DJ10,CJ11}. 

In all these situations, it remains to be seen whether some cosmological
variables might take values that differ from their canonical ones, as
derived in the framework of the best-fit vanilla single field inflation
paradigm. To clarify the situation, a full perturbation theory should
now be examined \cite{Zlosnik11}. Unlike those that assume only a single 
fluid, a perturbation analysis of a two-fluid system has to take into 
account the possibility of two-stream instability \cite{SLAC10}. 

Such instabilities are well-established in plasma physics, and have also been 
argued for in laboratory superfluids and their neutron star analogs 
\cite{andersson04:_twostream,acp03:_twostream_prl}. Samuelsson \etal 
\cite{SLAC10} have shown that a relative velocity and some type of coupling 
(cross-constituent or entrainment) for a system of two relativistic fluids 
is a necessary condition for two-stream instability. Roughly, there is a 
``window'' of instability that opens when a mode appears to be, say, 
right-moving with respect to one of the fluids, yet left-moving with 
respect to the other. In this paper we have shown that the conditions for 
such instabilities to exist, a relative flow between two coupled fluid components, 
may be satisfied in cosmology. We have also shown that cosmological 
expansion provides a mechanism for shutting down the instability by closing the 
window, since both the relative velocity and the cross-constituent coupling are 
driven to zero.

If such instabilities were to be triggered, a basis for a set of observational 
constraints (or possible detections) for the transition epoch may be 
established. In a companion paper \cite{cpaprl}, we explore whether these 
instabilities develop before the instability window is closed. If an 
instability were to develop in some of the anisotropic transitions, they 
would most definitely leave relevant imprints in both CMB and large scale 
structure data, in the form either of non-gaussianities, bizarre polarization 
distributions and spectra, and special scales corresponding to the Hubble 
volume at the transition time. For instance, it can be argued that such 
instabilities can occur during the matter to cosmological constant transition 
if and only if the latter is made of a fluid, hence having a state parameter 
$w>-1$; however close $w$ is to $-1$, such a fluid could initiate an 
instability that an actual cosmological constant, having $w=-1$, could not. 
Therefore, observing the relevant consequences of these instabilities at 
the relevant length scales would allow a discrimination between these two 
otherwise indistinguishable models. 

\acknowledgments

GLC acknowledges support from NSF via grant number PHYS-0855558. 
PP would like to thank support from the Perimeter Institute in which the
final part of this 
work has been done. NA acknowledges support from STFC in the UK.

\section*{Appendix: Geometric quantities}

For the metric given in (\eqref{fin_metric}) we find the Christoffel
coefficients to be
\bea
    \Gamma^t_{xx} &=& A^2_x H_x \quad , \quad \Gamma^t_{yy} = A^2_y H_y
                      \quad , \quad \Gamma^t_{zz} = A^2_z H_z , \cr
                   && \cr
    \Gamma^x_{tx} &=& H_x \quad , \quad
    \Gamma^2_{ty} = H_y \quad , \quad \Gamma^z_{tz} = H_z , \cr
                   && \cr
    \Gamma^x_{xz} &=& I_x \quad , \quad \Gamma^y_{yz} = I_y \quad , \quad
                      \Gamma^z_{zz} = I_z , \cr
                   && \cr
    \Gamma^z_{xx} &=& - \left(\frac{A_x}{A_z}\right)^2 I_x \quad , \quad
    \Gamma^z_{yy} = - \left(\frac{A_y}{A_z}\right)^2 I_y ,
\eea
leading to the following non-vanishing components of the Ricci tensor:
\begin{widetext}
\bea
R^x_{\ x} &=& H_x\sum_\aleph H_\aleph + H_x'-\frac{1}{A_z^2} \left[
  I'_x + I_x\left(I_x+I_y-I_z\right)\right] ,\\
R^y_{\ y} &=& H_y\sum_\aleph H_\aleph + H_y'-\frac{1}{A_z^2} \left[
  I'_y + I_y\left(I_x+I_y-I_z\right) \right] , \\
R^z_{\ z} &=& H_z\sum_\aleph H_\aleph + H_z'-\frac{1}{A_z^2} \left[
  I'_x + I'_y + I_x\left(I_x-I_z\right) + I_y \left( I_y - I_z\right)\right] 
              , \\
R^z_{\ t} &=& \frac{1}{A_z^2} \left[ I_x\left(H_z-H_x\right) + I_y
  \left(H_z-H_y\right) -I'_x - I'_y\right] , \\
R^t_{\ t} &=&\sum_\aleph \left( H_\aleph'+H_\aleph^2\right) ,
\label{Ricci}
\eea
and scalar
\beq
R = 2\left\{ \sum_\aleph \left( H_\aleph^2+H_\aleph'\right)+H_x H_y +
H_x H_x +H_y H_z -\frac{1}{A_z^2}\left[I_x'+I_y'+\left(I_x+I_y\right)^2 - 
I_x I_z - I_y I_z - I_x I_y \right] \right\} ,
\label{RicScalar}
\eeq
\end{widetext}
with the sign convention for the Riemann tensor given by
\beq
R^\mu_{\ \nu\alpha\beta} \equiv \partial_\beta\Gamma^\mu_{\nu\alpha}
- \partial_\alpha\Gamma^\mu_{\nu\beta} + \Gamma^\mu_{\sigma\beta}
\Gamma^\sigma_{\nu\alpha} -\Gamma^\mu_{\sigma\alpha}
\Gamma^\sigma_{\nu\beta} .
\label{SignRiemann}
\eeq
From these, one can obtain the Einstein tensor (\ref{eineqn}). As pointed 
out in the main text, the Einstein equations are not all independent.

\bibliography{Bib2Streams}

\begin{thebibliography}{53}
\expandafter\ifx\csname natexlab\endcsname\relax\def\natexlab#1{#1}\fi
\expandafter\ifx\csname bibnamefont\endcsname\relax
  \def\bibnamefont#1{#1}\fi
\expandafter\ifx\csname bibfnamefont\endcsname\relax
  \def\bibfnamefont#1{#1}\fi
\expandafter\ifx\csname citenamefont\endcsname\relax
  \def\citenamefont#1{#1}\fi
\expandafter\ifx\csname url\endcsname\relax
  \def\url#1{\texttt{#1}}\fi
\expandafter\ifx\csname urlprefix\endcsname\relax\def\urlprefix{URL }\fi
\providecommand{\bibinfo}[2]{#2}
\providecommand{\eprint}[2][]{\url{#2}}

\bibitem[{\citenamefont{{Peter} and {Uzan}}(2009)}]{PPJPU}
\bibinfo{author}{\bibfnamefont{P.}~\bibnamefont{{Peter}}} \bibnamefont{and}
  \bibinfo{author}{\bibfnamefont{J.-P.} \bibnamefont{{Uzan}}},
  \emph{\bibinfo{title}{Primordial cosmology}} (\bibinfo{publisher}{Oxford
  Graduate Texts, Oxford University press, UK}, \bibinfo{year}{2009}).

\bibitem[{\citenamefont{{Modak}}(1984)}]{modak}
\bibinfo{author}{\bibfnamefont{B.}~\bibnamefont{{Modak}}}, \bibinfo{journal}{J.
  Astrop. Astr.} \textbf{\bibinfo{volume}{5}}, \bibinfo{pages}{317}
  (\bibinfo{year}{1984}).

\bibitem[{\citenamefont{{Triginer} and {Pavon}}(1995)}]{pavon}
\bibinfo{author}{\bibfnamefont{J.}~\bibnamefont{{Triginer}}} \bibnamefont{and}
  \bibinfo{author}{\bibfnamefont{D.}~\bibnamefont{{Pavon}}},
  \bibinfo{journal}{Class.~Quant.~Grav.} \textbf{\bibinfo{volume}{12}},
  \bibinfo{pages}{689} (\bibinfo{year}{1995}).

\bibitem[{\citenamefont{{Andersson} and {Lopez-Monsalvo}}(2011)}]{AL11}
\bibinfo{author}{\bibfnamefont{N.}~\bibnamefont{{Andersson}}} \bibnamefont{and}
  \bibinfo{author}{\bibfnamefont{C.~S.} \bibnamefont{{Lopez-Monsalvo}}},
  \bibinfo{journal}{Classical and Quantum Gravity}
  \textbf{\bibinfo{volume}{28}}, \bibinfo{pages}{195023}
  (\bibinfo{year}{2011}).

\bibitem[{\citenamefont{{Weinberg}}(1971)}]{Weinberg71}
\bibinfo{author}{\bibfnamefont{S.}~\bibnamefont{{Weinberg}}},
  \bibinfo{journal}{\apj} \textbf{\bibinfo{volume}{168}}, \bibinfo{pages}{175}
  (\bibinfo{year}{1971}).

\bibitem[{\citenamefont{{Patel} and {Koppar}}(1991)}]{PK91}
\bibinfo{author}{\bibfnamefont{L.~K.} \bibnamefont{{Patel}}} \bibnamefont{and}
  \bibinfo{author}{\bibfnamefont{S.~S.} \bibnamefont{{Koppar}}},
  \bibinfo{journal}{Australian Mathematical Society Journal Series B -- Applied
  Mathematics} \textbf{\bibinfo{volume}{33}}, \bibinfo{pages}{77}
  (\bibinfo{year}{1991}).

\bibitem[{\citenamefont{Velten and Schwarz}(2011)}]{Velten:2011bg}
\bibinfo{author}{\bibfnamefont{H.}~\bibnamefont{Velten}} \bibnamefont{and}
  \bibinfo{author}{\bibfnamefont{D.~J.} \bibnamefont{Schwarz}},
  \bibinfo{journal}{JCAP} \textbf{\bibinfo{volume}{1109}}, \bibinfo{pages}{016}
  (\bibinfo{year}{2011}).

\bibitem[{\citenamefont{{Sikivie} and {Yang}}(2009)}]{SY09}
\bibinfo{author}{\bibfnamefont{P.}~\bibnamefont{{Sikivie}}} \bibnamefont{and}
  \bibinfo{author}{\bibfnamefont{Q.}~\bibnamefont{{Yang}}},
  \bibinfo{journal}{Physical Review Letters} \textbf{\bibinfo{volume}{103}},
  \bibinfo{pages}{111301} (\bibinfo{year}{2009}).

\bibitem[{\citenamefont{{Harko}}(2011)}]{Harko11}
\bibinfo{author}{\bibfnamefont{T.}~\bibnamefont{{Harko}}},
  \bibinfo{journal}{\prd} \textbf{\bibinfo{volume}{83}},
  \bibinfo{pages}{123515} (\bibinfo{year}{2011}).

\bibitem[{\citenamefont{Carter}(1989)}]{carter89:_covar_theor_conduc}
\bibinfo{author}{\bibfnamefont{B.}~\bibnamefont{Carter}}, in
  \emph{\bibinfo{booktitle}{Relativistic Fluid Dynamics (Noto, 1987)}}, edited
  by \bibinfo{editor}{\bibfnamefont{A.}~\bibnamefont{Anile}} \bibnamefont{and}
  \bibinfo{editor}{\bibfnamefont{M.}~\bibnamefont{Choquet-Bruhat}}
  (\bibinfo{publisher}{Springer-Verlag}, \bibinfo{address}{Heidelberg,
  Germany}, \bibinfo{year}{1989}), vol. \bibinfo{volume}{1385} of
  \emph{\bibinfo{series}{Lect. Notes Math.}}, pp. \bibinfo{pages}{1--64}.

\bibitem[{\citenamefont{{Andersson} and {Comer}}(2007)}]{andersson07:_livrev}
\bibinfo{author}{\bibfnamefont{N.}~\bibnamefont{{Andersson}}} \bibnamefont{and}
  \bibinfo{author}{\bibfnamefont{G.~L.} \bibnamefont{{Comer}}},
  \bibinfo{journal}{Living Reviews in Relativity}
  \textbf{\bibinfo{volume}{10}}, \bibinfo{pages}{1} (\bibinfo{year}{2007}).

\bibitem[{\citenamefont{{Nakar} et~al.}(2011)\citenamefont{{Nakar}, {Bret}, and
  {Milosavljevi{\'c}}}}]{NBM11}
\bibinfo{author}{\bibfnamefont{E.}~\bibnamefont{{Nakar}}},
  \bibinfo{author}{\bibfnamefont{A.}~\bibnamefont{{Bret}}}, \bibnamefont{and}
  \bibinfo{author}{\bibfnamefont{M.}~\bibnamefont{{Milosavljevi{\'c}}}},
  \bibinfo{journal}{\apj} \textbf{\bibinfo{volume}{738}}, \bibinfo{pages}{93}
  (\bibinfo{year}{2011}).

\bibitem[{\citenamefont{Gromov et~al.}(2004)\citenamefont{Gromov, Baryshev, and
  Teerikorpi}}]{Gromov:2002ek}
\bibinfo{author}{\bibfnamefont{A.}~\bibnamefont{Gromov}},
  \bibinfo{author}{\bibfnamefont{Y.}~\bibnamefont{Baryshev}}, \bibnamefont{and}
  \bibinfo{author}{\bibfnamefont{P.}~\bibnamefont{Teerikorpi}},
  \bibinfo{journal}{Astron. Astrophys.} \textbf{\bibinfo{volume}{415}},
  \bibinfo{pages}{813} (\bibinfo{year}{2004}).

\bibitem[{\citenamefont{{Emir G{\"u}mr{\"u}k{\c c}{\"u}oglu}
  et~al.}(2007)\citenamefont{{Emir G{\"u}mr{\"u}k{\c c}{\"u}oglu}, {Contaldi},
  and {Peloso}}}]{EGcCP07}
\bibinfo{author}{\bibfnamefont{A.}~\bibnamefont{{Emir G{\"u}mr{\"u}k{\c
  c}{\"u}oglu}}}, \bibinfo{author}{\bibfnamefont{C.~R.}
  \bibnamefont{{Contaldi}}}, \bibnamefont{and}
  \bibinfo{author}{\bibfnamefont{M.}~\bibnamefont{{Peloso}}},
  \bibinfo{journal}{JCAP} \textbf{\bibinfo{volume}{11}}, \bibinfo{pages}{5}
  (\bibinfo{year}{2007}).

\bibitem[{\citenamefont{{Pitrou} et~al.}(2008)\citenamefont{{Pitrou},
  {Pereira}, and {Uzan}}}]{PPU08}
\bibinfo{author}{\bibfnamefont{C.}~\bibnamefont{{Pitrou}}},
  \bibinfo{author}{\bibfnamefont{T.~S.} \bibnamefont{{Pereira}}},
  \bibnamefont{and} \bibinfo{author}{\bibfnamefont{J.-P.}
  \bibnamefont{{Uzan}}}, \bibinfo{journal}{JCAP} \textbf{\bibinfo{volume}{4}},
  \bibinfo{pages}{4} (\bibinfo{year}{2008}).

\bibitem[{\citenamefont{{Kim} and {Minamitsuji}}(2010)}]{KM10}
\bibinfo{author}{\bibfnamefont{H.-C.} \bibnamefont{{Kim}}} \bibnamefont{and}
  \bibinfo{author}{\bibfnamefont{M.}~\bibnamefont{{Minamitsuji}}},
  \bibinfo{journal}{\prd} \textbf{\bibinfo{volume}{81}},
  \bibinfo{pages}{083517} (\bibinfo{year}{2010}).

\bibitem[{\citenamefont{{Dechant} et~al.}(2009)\citenamefont{{Dechant},
  {Lasenby}, and {Hobson}}}]{DLH09}
\bibinfo{author}{\bibfnamefont{P.-P.} \bibnamefont{{Dechant}}},
  \bibinfo{author}{\bibfnamefont{A.~N.} \bibnamefont{{Lasenby}}},
  \bibnamefont{and} \bibinfo{author}{\bibfnamefont{M.~P.}
  \bibnamefont{{Hobson}}}, \bibinfo{journal}{\prd}
  \textbf{\bibinfo{volume}{79}}, \bibinfo{pages}{043524}
  (\bibinfo{year}{2009}).

\bibitem[{\citenamefont{{Dey} and {Paban}}(2011)}]{DP11}
\bibinfo{author}{\bibfnamefont{A.}~\bibnamefont{{Dey}}} \bibnamefont{and}
  \bibinfo{author}{\bibfnamefont{S.}~\bibnamefont{{Paban}}},
  \bibinfo{journal}{ArXiv e-prints}  (\bibinfo{year}{2011}).

\bibitem[{\citenamefont{{Sandin}}(2009)}]{Sandin09}
\bibinfo{author}{\bibfnamefont{P.}~\bibnamefont{{Sandin}}},
  \bibinfo{journal}{General Relativity and Gravitation}
  \textbf{\bibinfo{volume}{41}}, \bibinfo{pages}{2707} (\bibinfo{year}{2009}).

\bibitem[{\citenamefont{{Harko} and {Lobo}}(2011)}]{HL11}
\bibinfo{author}{\bibfnamefont{T.}~\bibnamefont{{Harko}}} \bibnamefont{and}
  \bibinfo{author}{\bibfnamefont{F.~S.~N.} \bibnamefont{{Lobo}}},
  \bibinfo{journal}{\prd} \textbf{\bibinfo{volume}{83}},
  \bibinfo{pages}{124051} (\bibinfo{year}{2011}).

\bibitem[{\citenamefont{{Calogero} and {Heinzle}}(2011)}]{CH10}
\bibinfo{author}{\bibfnamefont{S.}~\bibnamefont{{Calogero}}} \bibnamefont{and}
  \bibinfo{author}{\bibfnamefont{J.~M.} \bibnamefont{{Heinzle}}},
  \bibinfo{journal}{Physica} \textbf{\bibinfo{volume}{240}},
  \bibinfo{pages}{636} (\bibinfo{year}{2011}).

\bibitem[{\citenamefont{{Tsagas} et~al.}(2008)\citenamefont{{Tsagas},
  {Challinor}, and {Maartens}}}]{TCM08}
\bibinfo{author}{\bibfnamefont{C.~G.} \bibnamefont{{Tsagas}}},
  \bibinfo{author}{\bibfnamefont{A.}~\bibnamefont{{Challinor}}},
  \bibnamefont{and}
  \bibinfo{author}{\bibfnamefont{R.}~\bibnamefont{{Maartens}}},
  \bibinfo{journal}{Phys. Rep.} \textbf{\bibinfo{volume}{465}},
  \bibinfo{pages}{61} (\bibinfo{year}{2008}).

\bibitem[{\citenamefont{{Barrow} and {Tsagas}}(2007)}]{BT07}
\bibinfo{author}{\bibfnamefont{J.~D.} \bibnamefont{{Barrow}}} \bibnamefont{and}
  \bibinfo{author}{\bibfnamefont{C.~G.} \bibnamefont{{Tsagas}}},
  \bibinfo{journal}{Classical and Quantum Gravity}
  \textbf{\bibinfo{volume}{24}}, \bibinfo{pages}{1023} (\bibinfo{year}{2007}).

\bibitem[{\citenamefont{{Adhav} et~al.}(2011)\citenamefont{{Adhav}, {Borikar},
  {Desale}, and {Raut}}}]{ABDR11}
\bibinfo{author}{\bibfnamefont{K.~S.} \bibnamefont{{Adhav}}},
  \bibinfo{author}{\bibfnamefont{S.~M.} \bibnamefont{{Borikar}}},
  \bibinfo{author}{\bibfnamefont{M.~S.} \bibnamefont{{Desale}}},
  \bibnamefont{and} \bibinfo{author}{\bibfnamefont{R.~B.}
  \bibnamefont{{Raut}}}, \bibinfo{journal}{EJTP} \textbf{\bibinfo{volume}{8}},
  \bibinfo{pages}{319} (\bibinfo{year}{2011}).

\bibitem[{\citenamefont{{Cataldo} et~al.}(2011)\citenamefont{{Cataldo},
  {Ar\'evalo}, and {Mella}}}]{ACM11}
\bibinfo{author}{\bibfnamefont{M.}~\bibnamefont{{Cataldo}}},
  \bibinfo{author}{\bibfnamefont{F.}~\bibnamefont{{Ar\'evalo}}},
  \bibnamefont{and} \bibinfo{author}{\bibfnamefont{P.}~\bibnamefont{{Mella}}},
  \bibinfo{journal}{Astrophys. Space Sci.} \textbf{\bibinfo{volume}{333}},
  \bibinfo{pages}{287} (\bibinfo{year}{2011}).

\bibitem[{\citenamefont{{Schwarz} et~al.}(2004)\citenamefont{{Schwarz},
  {Starkman}, {Huterer}, and {Copi}}}]{SSHC04}
\bibinfo{author}{\bibfnamefont{D.~J.} \bibnamefont{{Schwarz}}},
  \bibinfo{author}{\bibfnamefont{G.~D.} \bibnamefont{{Starkman}}},
  \bibinfo{author}{\bibfnamefont{D.}~\bibnamefont{{Huterer}}},
  \bibnamefont{and} \bibinfo{author}{\bibfnamefont{C.~J.}
  \bibnamefont{{Copi}}}, \bibinfo{journal}{Physical Review Letters}
  \textbf{\bibinfo{volume}{93}}, \bibinfo{pages}{221301}
  (\bibinfo{year}{2004}).

\bibitem[{\citenamefont{{Copi} et~al.}(2010)\citenamefont{{Copi}, {Huterer},
  {Schwarz}, and {Starkman}}}]{CHSS10}
\bibinfo{author}{\bibfnamefont{C.~J.} \bibnamefont{{Copi}}},
  \bibinfo{author}{\bibfnamefont{D.}~\bibnamefont{{Huterer}}},
  \bibinfo{author}{\bibfnamefont{D.~J.} \bibnamefont{{Schwarz}}},
  \bibnamefont{and} \bibinfo{author}{\bibfnamefont{G.~D.}
  \bibnamefont{{Starkman}}}, \bibinfo{journal}{Advances in Astronomy}
  \textbf{\bibinfo{volume}{2010}}, \bibinfo{pages}{847541}
  (\bibinfo{year}{2010}).

\bibitem[{\citenamefont{{Perivolaropoulos}}(2011)}]{Perivolaropoulos11}
\bibinfo{author}{\bibfnamefont{L.}~\bibnamefont{{Perivolaropoulos}}},
  \bibinfo{journal}{ArXiv e-prints}  (\bibinfo{year}{2011}).

\bibitem[{\citenamefont{{Ma} et~al.}(2011)\citenamefont{{Ma}, {Efstathiou}, and
  {Challinor}}}]{MEC11}
\bibinfo{author}{\bibfnamefont{Y.-Z.} \bibnamefont{{Ma}}},
  \bibinfo{author}{\bibfnamefont{G.}~\bibnamefont{{Efstathiou}}},
  \bibnamefont{and}
  \bibinfo{author}{\bibfnamefont{A.}~\bibnamefont{{Challinor}}},
  \bibinfo{journal}{\prd} \textbf{\bibinfo{volume}{83}},
  \bibinfo{pages}{083005} (\bibinfo{year}{2011}).

\bibitem[{\citenamefont{{Pontzen} and {Challinor}}(2007)}]{PC07}
\bibinfo{author}{\bibfnamefont{A.}~\bibnamefont{{Pontzen}}} \bibnamefont{and}
  \bibinfo{author}{\bibfnamefont{A.}~\bibnamefont{{Challinor}}},
  \bibinfo{journal}{MNRAS} \textbf{\bibinfo{volume}{380}},
  \bibinfo{pages}{1387} (\bibinfo{year}{2007}).

\bibitem[{\citenamefont{{Pontzen}}(2009)}]{Pontzen09}
\bibinfo{author}{\bibfnamefont{A.}~\bibnamefont{{Pontzen}}},
  \bibinfo{journal}{\prd} \textbf{\bibinfo{volume}{79}},
  \bibinfo{pages}{103518} (\bibinfo{year}{2009}).

\bibitem[{\citenamefont{{Fixsen} and {Kashlinsky}}(2011)}]{FK11}
\bibinfo{author}{\bibfnamefont{D.~J.} \bibnamefont{{Fixsen}}} \bibnamefont{and}
  \bibinfo{author}{\bibfnamefont{A.}~\bibnamefont{{Kashlinsky}}},
  \bibinfo{journal}{\apj} \textbf{\bibinfo{volume}{734}}, \bibinfo{pages}{61}
  (\bibinfo{year}{2011}).

\bibitem[{\citenamefont{Komatsu et~al.}(2011)}]{Komatsu:2010fb}
\bibinfo{author}{\bibfnamefont{E.}~\bibnamefont{Komatsu}} \bibnamefont{et~al.},
  \bibinfo{journal}{Astrophys. J. Suppl.} \textbf{\bibinfo{volume}{192}},
  \bibinfo{pages}{18} (\bibinfo{year}{2011}).

\bibitem[{\citenamefont{{Percival} et~al.}(2010)\citenamefont{{Percival},
  {Reid}, {Eisenstein}, {Bahcall}, {Budavari}, {Frieman}, {Fukugita}, {Gunn},
  {Ivezi{\'c}}, {Knapp} et~al.}}]{Percival+10}
\bibinfo{author}{\bibfnamefont{W.~J.} \bibnamefont{{Percival}}},
  \bibinfo{author}{\bibfnamefont{B.~A.} \bibnamefont{{Reid}}},
  \bibinfo{author}{\bibfnamefont{D.~J.} \bibnamefont{{Eisenstein}}},
  \bibinfo{author}{\bibfnamefont{N.~A.} \bibnamefont{{Bahcall}}},
  \bibinfo{author}{\bibfnamefont{T.}~\bibnamefont{{Budavari}}},
  \bibinfo{author}{\bibfnamefont{J.~A.} \bibnamefont{{Frieman}}},
  \bibinfo{author}{\bibfnamefont{M.}~\bibnamefont{{Fukugita}}},
  \bibinfo{author}{\bibfnamefont{J.~E.} \bibnamefont{{Gunn}}},
  \bibinfo{author}{\bibfnamefont{{\v Z}.}~\bibnamefont{{Ivezi{\'c}}}},
  \bibinfo{author}{\bibfnamefont{G.~R.} \bibnamefont{{Knapp}}},
  \bibnamefont{et~al.}, \bibinfo{journal}{MNRAS}
  \textbf{\bibinfo{volume}{401}}, \bibinfo{pages}{2148} (\bibinfo{year}{2010}).

\bibitem[{\citenamefont{{Samuelsson} et~al.}(2010)\citenamefont{{Samuelsson},
  {Lopez-Monsalvo}, {Andersson}, and {Comer}}}]{SLAC10}
\bibinfo{author}{\bibfnamefont{L.}~\bibnamefont{{Samuelsson}}},
  \bibinfo{author}{\bibfnamefont{C.~S.} \bibnamefont{{Lopez-Monsalvo}}},
  \bibinfo{author}{\bibfnamefont{N.}~\bibnamefont{{Andersson}}},
  \bibnamefont{and} \bibinfo{author}{\bibfnamefont{G.~L.}
  \bibnamefont{{Comer}}}, \bibinfo{journal}{General Relativity and Gravitation}
  \textbf{\bibinfo{volume}{42}}, \bibinfo{pages}{413} (\bibinfo{year}{2010}).

\bibitem[{\citenamefont{{Comer} et~al.}(2011)\citenamefont{{Comer}, {Peter},
  and {Andersson}}}]{cpaprl}
\bibinfo{author}{\bibfnamefont{G.~L.} \bibnamefont{{Comer}}},
  \bibinfo{author}{\bibfnamefont{P.}~\bibnamefont{{Peter}}}, \bibnamefont{and}
  \bibinfo{author}{\bibfnamefont{N.}~\bibnamefont{{Andersson}}},
  \bibinfo{journal}{to be submitted}  (\bibinfo{year}{2011}).

\bibitem[{\citenamefont{{Comer} and {Joynt}}(2003)}]{comer03:_rel_ent}
\bibinfo{author}{\bibfnamefont{G.~L.} \bibnamefont{{Comer}}} \bibnamefont{and}
  \bibinfo{author}{\bibfnamefont{R.}~\bibnamefont{{Joynt}}},
  \bibinfo{journal}{Phys.~Rev.~D} \textbf{\bibinfo{volume}{68}},
  \bibinfo{pages}{023002} (\bibinfo{year}{2003}).

\bibitem[{\citenamefont{{Lemoine} et~al.}(2008)\citenamefont{{Lemoine},
  {Martin}, and {Peter}}}]{LNP2008}
\bibinfo{editor}{\bibfnamefont{M.}~\bibnamefont{{Lemoine}}},
  \bibinfo{editor}{\bibfnamefont{J.}~\bibnamefont{{Martin}}}, \bibnamefont{and}
  \bibinfo{editor}{\bibfnamefont{P.}~\bibnamefont{{Peter}}}, eds.,
  \emph{\bibinfo{title}{Inflationary Cosmology}}, vol. \bibinfo{volume}{738}
  (\bibinfo{year}{2008}).

\bibitem[{\citenamefont{{Ando} and {Kusenko}}(2010)}]{AK10}
\bibinfo{author}{\bibfnamefont{S.}~\bibnamefont{{Ando}}} \bibnamefont{and}
  \bibinfo{author}{\bibfnamefont{A.}~\bibnamefont{{Kusenko}}},
  \bibinfo{journal}{Ap. J. Lett.} \textbf{\bibinfo{volume}{722}},
  \bibinfo{pages}{L39} (\bibinfo{year}{2010}).

\bibitem[{\citenamefont{{Anber} and {Sorbo}}(2006)}]{AS06}
\bibinfo{author}{\bibfnamefont{M.~M.} \bibnamefont{{Anber}}} \bibnamefont{and}
  \bibinfo{author}{\bibfnamefont{L.}~\bibnamefont{{Sorbo}}},
  \bibinfo{journal}{JCAP} \textbf{\bibinfo{volume}{10}}, \bibinfo{pages}{18}
  (\bibinfo{year}{2006}).

\bibitem[{\citenamefont{{Watanabe} et~al.}(2009)\citenamefont{{Watanabe},
  {Kanno}, and {Soda}}}]{WKS09}
\bibinfo{author}{\bibfnamefont{M.-A.} \bibnamefont{{Watanabe}}},
  \bibinfo{author}{\bibfnamefont{S.}~\bibnamefont{{Kanno}}}, \bibnamefont{and}
  \bibinfo{author}{\bibfnamefont{J.}~\bibnamefont{{Soda}}},
  \bibinfo{journal}{Physical Review Letters} \textbf{\bibinfo{volume}{102}},
  \bibinfo{pages}{191302} (\bibinfo{year}{2009}).

\bibitem[{\citenamefont{{Watanabe} et~al.}(2011)\citenamefont{{Watanabe},
  {Kanno}, and {Soda}}}]{WKS11}
\bibinfo{author}{\bibfnamefont{M.-A.} \bibnamefont{{Watanabe}}},
  \bibinfo{author}{\bibfnamefont{S.}~\bibnamefont{{Kanno}}}, \bibnamefont{and}
  \bibinfo{author}{\bibfnamefont{J.}~\bibnamefont{{Soda}}},
  \bibinfo{journal}{MNRAS} \textbf{\bibinfo{volume}{412}}, \bibinfo{pages}{L83}
  (\bibinfo{year}{2011}).

\bibitem[{\citenamefont{{Wang} et~al.}(2011)\citenamefont{{Wang}, {Wu}, and
  {Hsu}}}]{WWH11}
\bibinfo{author}{\bibfnamefont{C.-H.} \bibnamefont{{Wang}}},
  \bibinfo{author}{\bibfnamefont{Y.-H.} \bibnamefont{{Wu}}}, \bibnamefont{and}
  \bibinfo{author}{\bibfnamefont{S.~D.~H.} \bibnamefont{{Hsu}}},
  \bibinfo{journal}{ArXiv e-prints}  (\bibinfo{year}{2011}).

\bibitem[{\citenamefont{{Christopherson}
  et~al.}(2009)\citenamefont{{Christopherson}, {Malik}, and
  {Matravers}}}]{CMM09}
\bibinfo{author}{\bibfnamefont{A.~J.} \bibnamefont{{Christopherson}}},
  \bibinfo{author}{\bibfnamefont{K.~A.} \bibnamefont{{Malik}}},
  \bibnamefont{and} \bibinfo{author}{\bibfnamefont{D.~R.}
  \bibnamefont{{Matravers}}}, \bibinfo{journal}{\prd}
  \textbf{\bibinfo{volume}{79}}, \bibinfo{pages}{123523}
  (\bibinfo{year}{2009}).

\bibitem[{\citenamefont{Ho\ifmmode~\check{r}\else
  \v{r}\fi{}ava}(2009)}]{HL2009}
\bibinfo{author}{\bibfnamefont{P.}~\bibnamefont{Ho\ifmmode~\check{r}\else
  \v{r}\fi{}ava}}, \bibinfo{journal}{Phys. Rev. D}
  \textbf{\bibinfo{volume}{79}}, \bibinfo{pages}{084008}
  (\bibinfo{year}{2009}).

\bibitem[{\citenamefont{{Blas} et~al.}(2010)\citenamefont{{Blas},
  {Pujol{\`a}s}, and {Sibiryakov}}}]{BPS10}
\bibinfo{author}{\bibfnamefont{D.}~\bibnamefont{{Blas}}},
  \bibinfo{author}{\bibfnamefont{O.}~\bibnamefont{{Pujol{\`a}s}}},
  \bibnamefont{and}
  \bibinfo{author}{\bibfnamefont{S.}~\bibnamefont{{Sibiryakov}}},
  \bibinfo{journal}{Phys. Rev. Lett.} \textbf{\bibinfo{volume}{104}},
  \bibinfo{pages}{181302} (\bibinfo{year}{2010}).

\bibitem[{\citenamefont{Jacobson}(2007)}]{Jacobson:2008aj}
\bibinfo{author}{\bibfnamefont{T.}~\bibnamefont{Jacobson}},
  \bibinfo{journal}{PoS} \textbf{\bibinfo{volume}{QG-PH}}, \bibinfo{pages}{020}
  (\bibinfo{year}{2007}).

\bibitem[{\citenamefont{Jacobson}(2010)}]{Jacobson2010}
\bibinfo{author}{\bibfnamefont{T.}~\bibnamefont{Jacobson}},
  \bibinfo{journal}{Phys.Rev.} \textbf{\bibinfo{volume}{D81}},
  \bibinfo{pages}{101502} (\bibinfo{year}{2010}).

\bibitem[{\citenamefont{{Donnelly} and {Jacobson}}(2010)}]{DJ10}
\bibinfo{author}{\bibfnamefont{W.}~\bibnamefont{{Donnelly}}} \bibnamefont{and}
  \bibinfo{author}{\bibfnamefont{T.}~\bibnamefont{{Jacobson}}},
  \bibinfo{journal}{\prd} \textbf{\bibinfo{volume}{82}},
  \bibinfo{pages}{064032} (\bibinfo{year}{2010}).

\bibitem[{\citenamefont{{Carruthers} and {Jacobson}}(2011)}]{CJ11}
\bibinfo{author}{\bibfnamefont{I.}~\bibnamefont{{Carruthers}}}
  \bibnamefont{and}
  \bibinfo{author}{\bibfnamefont{T.}~\bibnamefont{{Jacobson}}},
  \bibinfo{journal}{\prd} \textbf{\bibinfo{volume}{83}},
  \bibinfo{pages}{024034} (\bibinfo{year}{2011}).

\bibitem[{\citenamefont{{Zlosnik}}(2011)}]{Zlosnik11}
\bibinfo{author}{\bibfnamefont{T.~G.} \bibnamefont{{Zlosnik}}},
  \bibinfo{journal}{ArXiv e-prints}  (\bibinfo{year}{2011}).

\bibitem[{\citenamefont{{Andersson} et~al.}(2004)\citenamefont{{Andersson},
  {Comer}, and {Prix}}}]{andersson04:_twostream}
\bibinfo{author}{\bibfnamefont{N.}~\bibnamefont{{Andersson}}},
  \bibinfo{author}{\bibfnamefont{G.~L.} \bibnamefont{{Comer}}},
  \bibnamefont{and} \bibinfo{author}{\bibfnamefont{R.}~\bibnamefont{{Prix}}},
  \bibinfo{journal}{Mon.~Not.~R.~Astro.~Soc.} \textbf{\bibinfo{volume}{354}},
  \bibinfo{pages}{101} (\bibinfo{year}{2004}).

\bibitem[{\citenamefont{{Andersson} et~al.}(2003)\citenamefont{{Andersson},
  {Comer}, and {Prix}}}]{acp03:_twostream_prl}
\bibinfo{author}{\bibfnamefont{N.}~\bibnamefont{{Andersson}}},
  \bibinfo{author}{\bibfnamefont{G.~L.} \bibnamefont{{Comer}}},
  \bibnamefont{and} \bibinfo{author}{\bibfnamefont{R.}~\bibnamefont{{Prix}}},
  \bibinfo{journal}{Phys.~Rev.~Lett.} \textbf{\bibinfo{volume}{90}},
  \bibinfo{pages}{091101} (\bibinfo{year}{2003}).

\end{thebibliography}

\end{document}